\documentclass[aps, prl, twocolumn, longbibliography,superscriptaddress, floatfix,nofootinbib]{revtex4-2}
\usepackage[T1]{fontenc}
\usepackage[utf8]{inputenc}
\usepackage{amsmath}
\usepackage{amssymb}
\usepackage{graphicx}
\usepackage{xcolor}
\usepackage[normalem]{ulem}
\usepackage{microtype}
\usepackage{changes}



\usepackage{babel}
\begin{document}
\global\long\def\ket#1{\left| #1 \right\rangle }%
 
\global\long\def\mcH{{\mathcal{H}}}%
 
\global\long\def\lrp#1{\left( #1 \right)}%
 
\global\long\def\lrb#1{\left[ #1 \right]}%
 
\global\long\def\lrc#1{\left\{  #1 \right\}  }%
\global\long\def\bra#1{\left\langle #1\right|}%
\global\long\def\avg#1{\left\langle #1 \right\rangle }%

\global\long\def\hatU{\hat{U}}%
 
\global\long\def\cohU{\hat{U}^{\dagger}}%
\global\long\def\coa{\hat{a}^{\dagger}}%
 
\global\long\def\aoa{\hat{a}}%
 
\global\long\def\cob{\hat{b}^{\dagger}}%
 
\global\long\def\aob{\hat{b}}%
 
\global\long\def\coc{\hat{c}^{\dagger}}%
 
\global\long\def\aoc{\hat{c}}%
 
\global\long\def\cod{\hat{d}^{\dagger}}%
 
\global\long\def\aod{\hat{d}}%
 
\global\long\def\hatD{\hat{D}}%
 
\global\long\def\hatH{\hat{H}}%
\global\long\def\hatN{\hat{N}}%
\global\long\def\aoA{\hat{A}}%
 
\global\long\def\aoB{\hat{B}}%
\global\long\def\hatQ{\hat{Q}}%
\global\long\def\pdd#1#2{\frac{\partial#1}{\partial#2}}%
\global\long\def\hatrho{\hat{\rho}}%
 
\global\long\def\tdpsi{\tilde{\psi}}%
\global\long\def\hatO{\hat{O}}%
\global\long\def\aor{\hat{\rho}}%
\global\long\def\hIO{\hat{O}^{I}}%
\global\long\def\hIH{\hat{H}^{I}}%
\global\long\def\bpsi{\bar{\psi}}%
\global\long\def\bphi{\bar{\phi}}%
\global\long\def\bPsi{\bar{\Psi}}%
\global\long\def\bPhi{\bar{\Phi}}%
\global\long\def\bvpsi{\breve{\psi}}%
\global\long\def\bvphi{\breve{\phi}}%
\global\long\def\bvbphi{\breve{\bphi}}%
\global\long\def\bvbpsi{\breve{\bpsi}}%
\global\long\def\barn{\bar{n}}%
\global\long\def\barT{\bar{T}}%
\global\long\def\barc{\bar{c}}%
\global\long\def\bart{\bar{t}}%
\global\long\def\tdJ{\tilde{J}}%
\global\long\def\tdD{\tilde{D}}%
\global\long\def\barj{\bar{j}}%
\global\long\def\tdmcT{\mathcal{\tilde{T}}}%
\global\long\def\tdG{\tilde{G}}%

\global\long\def\bmeta{\boldsymbol{\eta}}%
 
\global\long\def\bmlm{{\bf \text{\ensuremath{\bm{\lambda}}}}}%
\global\long\def\bmxi{{\bf \text{\ensuremath{\bm{\xi}}}}}%
\global\long\def\bfd{{\bf d}}%
\global\long\def\bmk{\bm{k}}%
\global\long\def\bmn{\bm{n}}%
\global\long\def\bmq{\bm{q}}%
\global\long\def\bmx{\bm{x}}%
\global\long\def\bmy{\bm{y}}%

\global\long\def\bmA{\bm{A}}%
\global\long\def\bmE{\bm{E}}%
\global\long\def\bmI{{\bf I}}%
\global\long\def\bmJ{\bm{J}}%
\global\long\def\bmM{\bm{M}}%
\global\long\def\bmN{\bm{N}}%
\global\long\def\bmX{\bm{X}}%
\global\long\def\bmM{\bm{M}}%
\global\long\def\bmE{\bm{E}}%
\global\long\def\bmS{\bm{S}}%

\global\long\def\rmCAR{{\rm CAR}}%
\global\long\def\rmLAR{{\rm LAR}}%
\global\long\def\rmNT{{\rm NT}}%
\global\long\def\bmA{\bm{A}}%
\global\long\def\rmCT{{\rm CT}}%
\global\long\def\rmQD{{\rm QD}}%
\global\long\def\rmI{{\rm I}}%
\global\long\def\rmeff{{\rm eff}}%
\global\long\def\rmD{{\rm D}}%
\global\long\def\rmMZM{{\rm MZM}}%
\global\long\def\rmwith{{\rm with}}%
\global\long\def\rmtot{{\rm tot}}%

\global\long\def\tr{\mathrm{Tr}}%
\global\long\def\half{\mathrm{\frac{1}{2}}}%
\global\long\def\intw{{\bf \text{\ensuremath{\int\frac{d\omega}{2\pi}}}}}%

\global\long\def\mcA{{\mathcal{A}}}%
 
\global\long\def\mcB{{\mathcal{B}}}%
 
\global\long\def\mcC{{\mathcal{C}}}%
 
\global\long\def\mcD{{\mathcal{D}}}%
 
\global\long\def\mcE{{\mathcal{E}}}%
 
\global\long\def\mcF{{\mathcal{F}}}%
 
\global\long\def\mcG{{\mathcal{G}}}%
 
\global\long\def\mcH{{\mathcal{H}}}%
 
\global\long\def\mcI{{\mathcal{I}}}%
 
\global\long\def\mcJ{{\mathcal{J}}}%
 
\global\long\def\mcK{{\mathcal{K}}}%
 
\global\long\def\mcL{{\mathcal{L}}}%
 
\global\long\def\mcM{{\mathcal{M}}}%
 
\global\long\def\mcN{\mathcal{N}}%
 
\global\long\def\mcO{{\mathcal{O}}}%
 
\global\long\def\mcP{{\mathcal{P}}}%
 
\global\long\def\mcQ{{\mathcal{Q}}}%
 
\global\long\def\mcR{{\mathcal{R}}}%
 
\global\long\def\mcS{{\mathcal{S}}}%
 
\global\long\def\mcT{\mathcal{T}}%
 
\global\long\def\mcU{{\mathcal{U}}}%
 
\global\long\def\mcV{{\mathcal{V}}}%
 
\global\long\def\mcW{{\mathcal{W}}}%
 
\global\long\def\mcX{{\mathcal{X}}}%
 
\global\long\def\mcY{{\mathcal{Y}}}%
 
\global\long\def\mcZ{{\mathcal{Z}}}%

\global\long\def\bbA{{\mathbb{A}}}%
 
\global\long\def\bbB{{\mathbb{B}}}%
 
\global\long\def\bbC{{\mathbb{C}}}%
 
\global\long\def\bbD{{\mathbb{D}}}%
 
\global\long\def\bbE{{\mathbb{E}}}%
 
\global\long\def\bbF{{\mathbb{F}}}%
 
\global\long\def\bbG{{\mathbb{G}}}%
 
\global\long\def\bbH{{\mathbb{H}}}%
 
\global\long\def\bbI{{\mathbb{I}}}%
 
\global\long\def\bbJ{{\mathbb{J}}}%
 
\global\long\def\bbK{{\mathbb{K}}}%
 
\global\long\def\bbL{{\mathbb{L}}}%
 
\global\long\def\bbM{{\mathbb{M}}}%
 
\global\long\def\bbN{{\mathbb{N}}}%
 
\global\long\def\bbO{{\mathbb{O}}}%
 
\global\long\def\bbP{{\mathbb{P}}}%
 
\global\long\def\bbQ{{\mathbb{Q}}}%
 
\global\long\def\bbR{{\mathbb{R}}}%
 
\global\long\def\bbS{{\mathbb{S}}}%
 
\global\long\def\bbT{\mathbb{T}}%
 
\global\long\def\bbU{{\mathbb{U}}}%
 
\global\long\def\bbV{{\mathbb{V}}}%
 
\global\long\def\bbW{{\mathbb{W}}}%
 
\global\long\def\bbX{{\mathbb{X}}}%
 
\global\long\def\bbY{{\mathbb{Y}}}%
 
\global\long\def\bbZ{{\mathbb{Z}}}%

\global\long\def\mfa{{\mathfrak{a}}}%
 
\global\long\def\mfb{{\mathfrak{b}}}%
 
\global\long\def\mfc{{\mathfrak{c}}}%
 
\global\long\def\mfd{{\mathfrak{d}}}%
 
\global\long\def\mfe{{\mathfrak{e}}}%
 
\global\long\def\mff{{\mathfrak{f}}}%
 
\global\long\def\mfg{{\mathfrak{g}}}%
 
\global\long\def\mfh{{\mathfrak{h}}}%
 
\global\long\def\mfi{{\mathfrak{i}}}%
 
\global\long\def\mfj{{\mathfrak{j}}}%
 
\global\long\def\mfk{{\mathfrak{k}}}%
 
\global\long\def\mfl{{\mathfrak{l}}}%
 
\global\long\def\mfm{{\mathfrak{m}}}%
 
\global\long\def\mfn{{\mathfrak{n}}}%
 
\global\long\def\mfo{{\mathfrak{o}}}%
 
\global\long\def\mfp{{\mathfrak{p}}}%
 
\global\long\def\mfq{{\mathfrak{q}}}%
 
\global\long\def\mfr{{\mathfrak{r}}}%
 
\global\long\def\mfs{{\mathfrak{s}}}%
 
\global\long\def\mft{{\mathfrak{t}}}%
 
\global\long\def\mfu{{\mathfrak{u}}}%
 
\global\long\def\mfv{{\mathfrak{v}}}%
 
\global\long\def\mfw{{\mathfrak{w}}}%
 
\global\long\def\mfx{{\mathfrak{x}}}%
 
\global\long\def\mfy{{\mathfrak{y}}}%
 
\global\long\def\mfz{{\mathfrak{z}}}%

\global\long\def\mfA{{\mathfrak{A}}}%
 
\global\long\def\mfB{{\mathfrak{B}}}%
 
\global\long\def\mfC{{\mathfrak{C}}}%
 
\global\long\def\mfD{{\mathfrak{D}}}%
 
\global\long\def\mfE{{\mathfrak{E}}}%
 
\global\long\def\mfF{{\mathfrak{F}}}%
 
\global\long\def\mfG{{\mathfrak{G}}}%
 
\global\long\def\mfH{{\mathfrak{H}}}%
 
\global\long\def\mfI{{\mathfrak{I}}}%
 
\global\long\def\mfJ{{\mathfrak{J}}}%
 
\global\long\def\mfK{{\mathfrak{K}}}%
 
\global\long\def\mfL{{\mathfrak{L}}}%
 
\global\long\def\mfM{{\mathfrak{M}}}%
 
\global\long\def\mfN{{\mathfrak{N}}}%
 
\global\long\def\mfO{{\mathfrak{O}}}%
 
\global\long\def\mfP{{\mathfrak{P}}}%
 
\global\long\def\mfQ{{\mathfrak{Q}}}%
 
\global\long\def\mfR{{\mathfrak{R}}}%
 
\global\long\def\mfS{{\mathfrak{S}}}%
 
\global\long\def\mfT{{\mathfrak{T}}}%
 
\global\long\def\mfU{{\mathfrak{U}}}%
 
\global\long\def\mfV{{\mathfrak{V}}}%
 
\global\long\def\mfW{{\mathfrak{W}}}%
 
\global\long\def\mfX{{\mathfrak{X}}}%
 
\global\long\def\mfY{{\mathfrak{Y}}}%
 
\global\long\def\mfZ{{\mathfrak{Z}}}%

\global\long\def\mrA{{\mathrm{A}}}%
 
\global\long\def\mrB{{\mathrm{B}}}%
 
\global\long\def\mrC{{\mathrm{C}}}%
 
\global\long\def\mrD{{\mathrm{D}}}%
 
\global\long\def\mrE{{\mathrm{E}}}%
 
\global\long\def\mrF{{\mathrm{F}}}%
 
\global\long\def\mrG{{\mathrm{G}}}%
 
\global\long\def\mrH{{\mathrm{H}}}%
 
\global\long\def\mrI{{\mathrm{I}}}%
 
\global\long\def\mrJ{{\mathrm{J}}}%
 
\global\long\def\mrK{{\mathrm{K}}}%
 
\global\long\def\mrL{{\mathrm{L}}}%
 
\global\long\def\mrM{{\mathrm{M}}}%
 
\global\long\def\mrN{{\mathrm{N}}}%
 
\global\long\def\mrO{{\mathrm{O}}}%
 
\global\long\def\mrP{{\mathrm{P}}}%
 
\global\long\def\mrQ{{\mathrm{Q}}}%
 
\global\long\def\mrR{{\mathrm{R}}}%
 
\global\long\def\mrS{{\mathrm{S}}}%
 
\global\long\def\mrT{{\mathrm{T}}}%
 
\global\long\def\mrU{{\mathrm{U}}}%
 
\global\long\def\mrV{{\mathrm{V}}}%
 
\global\long\def\mrW{{\mathrm{W}}}%
 
\global\long\def\mrX{{\mathrm{X}}}%
 
\global\long\def\mrY{{\mathrm{Y}}}%
 
\global\long\def\mrZ{{\mathrm{Z}}}%

\global\long\def\msH{\mathscr{H}}%
\global\long\def\msK{\mathscr{K}}%
\global\long\def\msG{\mathscr{G}}%
 
\global\long\def\tdbbT{\tilde{\bbT}}%
\newcommand{\del}[1]{\textcolor{red}{\sout{#1}}}
\newcommand{\issue}[1]{\textcolor{red}{#1}}

\preprint{APS/123-QED}
\title{Criticality around the Spinodal Point of First-Order Quantum Phase Transitions}
\author{Fan Zhang}
\email{van314159@pku.edu.cn}
\affiliation{School of Physics, Peking University, Beijing, 100871, China}

\author{Chiao Wang}
\affiliation{School of Physics, Peking University, Beijing, 100871, China}

\author{H. T. Quan}
\email{htquan@pku.edu.cn}
\affiliation{School of Physics, Peking University, Beijing, 100871, China}
\affiliation{Collaborative Innovation Center of Quantum Matter, Beijing 100871,
China}
\affiliation{Frontiers Science Center for Nano-optoelectronics, Peking University,
Beijing, 100871, China}

\date{\today}
\begin{abstract}
Universality and scaling are hallmarks of second-order phase transitions but are generally unexpected in first-order quantum phase transitions (FOQPTs). 
We present a microscopic theory showing that quantum criticality can emerge around the \textit{quantum spinodal point} of FOQPTs where metastability disappears. We demonstrate that, at this instability, resonant local excitations dynamically decouple a Hilbert subspace characterized by an emergent discrete translational symmetry. Projecting the original Hamiltonian onto this subspace yields an effective Hamiltonian that exhibits a genuine second-order quantum phase transition (SOQPT) and the Kibble–Zurek scaling.
We validate this framework in the tilted Ising chain which breaks $\bbZ_2$ symmetry, and predict the absence of criticality in the staggered-field PXP model. This work indicates that the dynamics of FOQPTs is usually governed by an emergent critical point around the quantum spinodal point. Our results uncover a hidden criticality in FOQPTs, reshaping the conventional understanding of FOQPTs beyond the mean-field theory. 
%

\end{abstract}

\maketitle

\textit{Introduction.}--- Understanding nonequilibrium dynamics across phase transitions is a central problem in modern physics. For second-order phase transitions, universality and scaling, which originate from the divergence of the correlation length, are well understood within the renormalization-group framework~\cite{sachdev2011book}. When a system is driven across a critical point at a finite ramp rate $v$, these universal properties are extended to nonequilibrium dynamics through the Kibble-Zurek mechanism (KZM)~\cite{Kibble1976jpa,Zurek1985nature,Dziarmaga2010AP}, which predicts a power-law scaling of the density of topological defects with the ramp rate $v$. 


In contrast, first-order phase transitions (FOPTs) have traditionally been regarded as lacking critical behavior at the phase transition point. Instead, they are characterized by phase coexistence, metastability, ergodicity breaking, hysteresis, and nucleation~\cite{Langer1969AP,Binder1987RPP,zinn-justin2021,miyashita2022book,kamenev2023}. In the past decades, some efforts have been devoted to uncovering scaling properties associated with FOPTs. The mean-field theory reveals that a universal scaling behavior without symmetry breaking can emerge, not at the phase transition point, but around the \textit{spinodal point} where the metastability disappears~\cite{ikeda1979PoTP,gunton1978PRB,jung1990PRL,zhong2005PRL,Bapst2012JSP,zhong2017FP,wu2025prl}. In this case, the transition time and the ramp rate are found to satisfy the power-law scaling relation $t_\mathrm{trans} \sim v^{-1/3}$~\cite{zhong2005PRL,zhong2017FP, kundu2023PRE,wu2025prl,Chen2026prl}. Beyond the mean-field theory, almost no consensus on scaling behaviors has been reached~\cite{Binder1987RPP, Acharyya1995PRB,Chakrabarti1999RMP,Lo1999PRA,Scopa2018jsm, Zhang2025CPL,Pelissetto2026PRE,sun2026arxiv}. For example, in the 2D classical Ising model driven across a magnetic FOPT, the reported scaling exponents vary from one study to another~\cite{Sengupta1992PRB, Klein1983prb,zheng1998jpcm, Chakrabarti1999RMP,Lo1999PRA,Zhang2025CPL,Pelissetto2026PRE,sun2026arxiv, Sieke2026arxiv}. In another example, 
a recent large-scale simulation has reported a power-law scaling in a FOQPT model~\cite{Vodeb2025NP}, 
but some theoretical investigations predict an unconventional logarithmic spinodal-like behavior in the same model~\cite{Pelissetto2025prb,Pelissetto2025PRBb}. This complexity is partly due to the lack of well-defined concepts of metastability and spinodal point in realistic short-range models~\cite{saito1978PTP,Binder1987RPP,pelissetto2017PRLa,pelissetto2023,Pelissetto2026PRE}. Recently, Yin et al.~\cite{yin2025PRX} proposed a rigorous definition of metastability in short-range quantum systems. This definition provides a new tool for investigating the dynamics of FOQPTs beyond mean-field models. 
In this Letter, we develop a microscopic theory of FOQPT dynamics based on the notion of $(\Delta,R)$-metastability proposed in Ref.~\cite{yin2025PRX}. We define the quantum spinodal point at which $(\Delta,R)$-metastability disappears for the minimum $R$. Starting from this point, 
using the Schrieffer–Wolff transformation, we obtain an effective Hamiltonian and show that the resonant local excitations dynamically decouple a Hilbert subspace with an emergent discrete translational symmetry~(e.g., $\mathbb{Z}_n$). 
Within this subspace, the effective Hamiltonian exhibits a SOQPT that breaks the emergent discrete translational symmetry  absent from the original Hamiltonian. The disappearance of the FOQPT metastability can thus be mapped onto an effective SOQPT in the dynamically decoupled subspace, leading to genuine criticality and Kibble–Zurek scaling.

We demonstrate this mechanism in two paradigmatic models. The first model is the one-dimensional quantum tilted Ising chain~\cite{Vodeb2025NP, Pelissetto2025prb,Pelissetto2025PRBb}, which around the spinodal point is projected onto a PXP-type constrained model with an emergent $\mathbb{Z}_2$ symmetry. The second model is the PXP model with a staggered field (the FOQPT properties of this model are given in \cite{supp}). Here, the dynamical constraint of the PXP model precludes additional symmetry breaking, resulting in the absence of an effective SOQPT and of the universal scaling behaviors. In the first model, we test the Kibble-Zurek scaling and the finite-time scaling~\cite{gong2010NJP,Zhong2011book,supp} to quantitatively confirm that the dynamics is governed by the effective SOQPT. We also provide another model in \cite{supp}, where the emergent symmetry is $\mathbb{Z}_3$ and the criticality is observed. In the second model, we analytically and numerically demonstrate that the dynamics is reduced to many independent Landau-Zener transitions, which do not exhibit scaling behaviors. 
Our results demonstrate that the dynamical scaling in FOQPTs originates from an emergent quantum critical point embedded in the metastable sector near the quantum spinodal. This recasts the dynamics of FOQPTs through the lens of quantum criticality, challenging the conventional view that universal scaling is exclusive to SOPTs.

\begin{figure}
\includegraphics[scale=0.15]{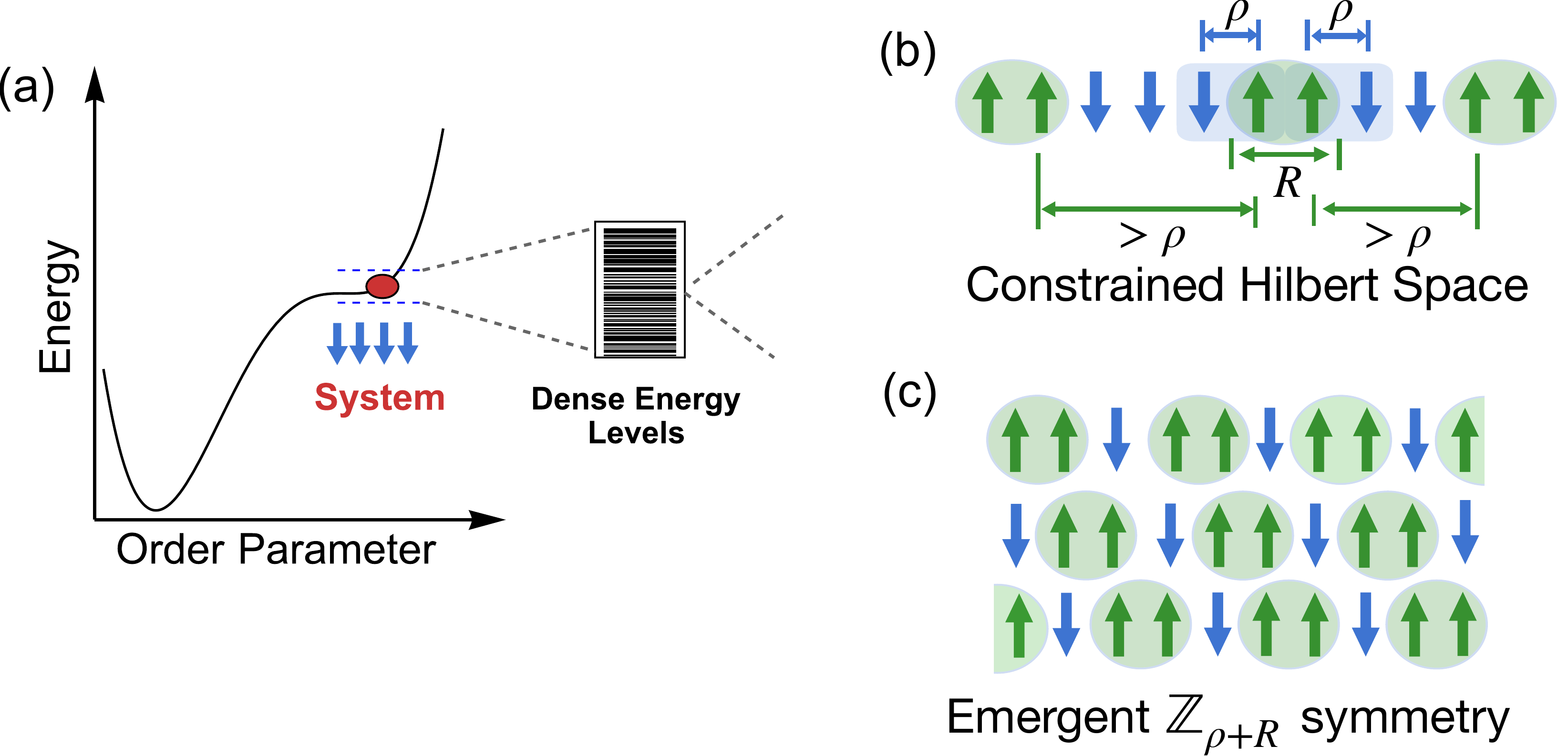}\caption{
Emergent symmetry at a quantum spinodal point.
(a) Schematic energy landscape near the quantum spinodal point. Taking the all-down state as the metastable state, the spinodal point is reached when a local excitation costs zero energy, giving rise to a dense set of degenerate energy levels.
(b) The degenerate states are formed by local excitations of size $R$ on top of the metastable background. Distinct excitations must be separated by at least $\rho+1$ where $\rho$ is the interaction range of the Hamiltonian. These states define a kinetically constrained Hilbert space.
(c) Within this constrained Hilbert space, the effective Hamiltonian near the quantum spinodal point exhibits an emergent $\mathbb{Z}_{\rho+R
}$ symmetry. For the example shown in (b) and (c), $R=2$ and $\rho=1$, giving an emergent $\mathbb{Z}_3$ symmetry. The three symmetry-related states are generated by cyclic translations of the local-excitation pattern.
}
\label{fig:ene_spec}
\end{figure}

\textit{General theory.}---We consider a local Hamiltonian $H(\lambda_t)$ which features a FOQPT, where $\lambda_t = vt$ is a time-dependent parameter that drives the system from one phase to another. Infinite-range models, such as the Lipkin-Meshkov-Glick model~\cite{lipin1965np} and quantum $p$-spin model~\cite{jorg2010EL}, are thus excluded; their dynamics are well captured by quantum mean-field theory~\cite{Bapst2012JSP}. We denote the interaction range of $H(\lambda_t)$ by $\rho$, defined as the number of sites in the largest support of any single term in the Hamiltonian. For example, $\sigma_j^z\sigma_{j+1}^z$ gives $\rho = 1$, and $\sigma_j^z\sigma_{j+2}^z$ gives $\rho = 2$ where $\sigma_j^z$ is the Pauli $z$ operator at site $j$.

To define the quantum spinodal point rigorously, we adapt the concept of $(\Delta, R)$-metastability from Ref.~\cite{yin2025PRX}. Given a local Hermitian operator $\mcO$ with the support diameter $l \leq R$ (e.g., $l = 2$ for $\mcO = \sigma_j^z\sigma_{j+1}^z$, $l=3$ for $\mcO = \sigma_j^z \sigma_{j+1}^z \sigma_{j+2}^z$), a state $\ket{\psi}$ is $(\Delta, R)$-metastable with respect to $H(\lambda_t)$ if, for every such $\mcO$, the perturbed state $(\mcO - \avg{\mcO}_{\psi})\ket{\psi} \not\propto \ket{\psi}$ raises the mean energy by at least $\Delta$:
\begin{equation}
\avg{H(\lambda_{t})}_{(\mcO-\avg{\mcO}_{\psi})\ket{\psi}}-\langle H(\lambda_{t})\rangle_{\ket{\psi}}\geq\Delta,
\label{eq:metastability}
\end{equation}
where $\avg{\mcO}_{\ket{\psi}} \equiv \bra{\psi}\mcO\ket{\psi}/\langle\psi|\psi\rangle$. Intuitively, $\ket{\psi}$ is a local minimum of the energy landscape with width $R$ and barrier height $\Delta$. Note that $\ket{\psi}$ need not be an eigenstate of $H(\lambda_t)$.

We define the \textit{quantum spinodal point} $\lambda^{\mathrm{sp}}$ as the parameter value at which the system loses $(\Delta, R)$-metastability for the minimal admissible $R$. Concretely, this is the threshold at which there exists a local operator $\mcO_j$ supported on a region of diameter $R$ centered at site $j$ such that the mean energy change vanishes:
\begin{equation}
\avg{H(\lambda^{\mathrm{sp}})}_{(\mcO_{j}-\avg{\mcO_{j}}_{\ket{\psi}})\ket{\psi}}-\langle H(\lambda^{\mathrm{sp}})\rangle_{\ket{\psi}}=0.
\label{eq:quan_spinodal_condition}
\end{equation}
Intuitively, $\mcO_j$ creates a local excitation on top of $\ket{\psi}$, and Eq.~\eqref{eq:quan_spinodal_condition} states that this excitation costs no energy at $\lambda^{\mathrm{sp}}$, see Fig.~\ref{fig:ene_spec}(a). 

Translating $\mcO_j$ across the lattice generates a family of zero-energy excitations. Multi-excitation states $\mcO_{j_1} \mcO_{j_2} \cdots \mcO_{j_n}\ket{\psi}$ remain degenerate with $\ket{\psi}$ provided that no term in $H(\lambda^{\mathrm{sp}})$ couples two distinct excitations. This means that any two operators $\{\mcO_j\}$ are at least separated by $\rho+1$. Configurations with closer-spaced excitations cost a finite energy and are gapped from the degenerate manifold. The resulting subspace is therefore an isolated subspace with dynamical constraints, analogous to Rydberg blockade models~\cite{bernien2017Nature, turner2018NP}, see Fig.~\ref{fig:ene_spec}(b).

This blockade structure imposes an emergent discrete translational symmetry $\mathbb{Z}_n$ in the constrained subspace, with periodicity
\begin{equation}
n = \rho + R,
\label{eq:period}
\end{equation}
set jointly by the interaction range of $H(\lambda^{\mathrm{sp}})$ and $R$ (see Fig.~\ref{fig:ene_spec}(c)). The effective Hamiltonian on this subspace, obtained by projecting $H(\lambda^{\mathrm{sp}})$ onto the manifold, is a generalized PXP-type model.

When the emergent $\mathbb{Z}_n$ symmetry is spontaneously broken during the driving process, the dynamics exhibits genuine quantum criticality and Kibble-Zurek scaling. Conversely, when constraints on the original Hilbert space preclude the symmetry breaking, no effective critical point appears, and no universal scaling is observed. This mechanism underlies the diverse dynamical behaviors seen in FOQPTs and provides a unified microscopic explanation for universality in their nonequilibrium response.


The effective Hamiltonian near the quantum spinodal point can be obtained by Schrieffer-Wolff transformation~\cite{bravyi2011AoP}, and the emergent criticality can be verified by the Kibble-Zurek scaling as well as the finite-time scaling~\cite{supp}. When the system is driven across the quantum spinodal point at a finite rate $v$, the density of topological defects $n$ at the final time satisfies the scaling relation
\begin{equation}
    n \sim v^{d\nu/(1+\nu z)},
\end{equation}
where $d$ is the spatial dimension of the system, $\nu$ is the critical exponent and $z$ is the dynamical critical exponent of the emergent critical point. Equivalently, one can examine the correlation length which obeys a finite-time scaling~\cite{gong2010NJP, Zhong2011book,huang2014PRB} 
\begin{equation}
\xi(\lambda,v) = v^{-\nu/(1+\nu z)} f\left( (\lambda-\lambda^{\mathrm{sp}}) v^{-1/(1+\nu z)} \right),
\end{equation}
where $f(x)$ is the non-universal scaling function for PXP model. These two relations will be used to quantitatively detect the presence of the effective SOQPT near the quantum spinodal point. In the following, we demonstrate our theory in two examples. A third example similar to the first example is given in the Supplemental Material~\cite{supp}.

\emph{Example 1: 1D tilted Ising model}.--- Our first model is the one-dimensional tilted quantum Ising model~(QIM) which is a prototype model of quantum phase transitions that exhibits both FOQPT and SOQPT. Its Hamiltonian reads
\begin{align}
H=-J\sum_j\sigma_{j}^{z}\sigma_{j+1}^{z}-h_{z}\sum_j\sigma_{j}^{z}-h_{x}\sum_j\sigma_{j}^{x},   \label{Eq:ham}
\end{align}
where $\sigma_{j}^{x,y,z}$ is the Pauli spin-$1/2$ matrix at site $j$, $h_{z}$ and $h_{x}$ are the longitudinal and transverse field, respectively. The ferromagnetic coupling $J$ is set to $1$ hereafter. This model is of paramount importance in various fields, from statistical mechanics and condensed matter to high-energy physics, and serves as a versatile platform to explore exotic phenomena, such as confinement, string breaking, bubble proliferation, and quantum chaos~\cite{sinha2021PRB, lagnese2021PRB, mirkin2021PRE, verdel2023PRL}. Further, it can be naturally realized in present-day Rydberg quantum simulators, and solid-state materials~\cite{kim2011NJP,lienhard2018PRX, lienhard2018PRX,liao2021PRA}.

The model (\ref{Eq:ham}) undergoes a ferromagnetic FOQPT at $h_{z}=0$ when $h_{x}<1$, and enters a paramagnetic phase when $h_{x}>1$ through the SOQPT. Its non-interacting version~($h_{z}=0$), the transverse QIM, is exactly solvable~\cite{dziarmaga2005PRL}. 
For the non-integrable tilted QIM, the dynamics is more involved. A recent study~\cite{sinha2021PRB} showed that, under slow tuning of the longitudinal field across the FOQPT line, bubbles of the true vacuum nucleated on top of the metastable false vacuum on a multitude of resonant points/regions. From the perspective of $(\Delta,R)$-metastability, these resonances correspond to the disappearance of metastability for different $R$, i.e., $\Delta=0$ at the resonant points. We will focus on the largest resonant point which corresponds to the minimum $R=1$ and thus the quantum spinodal point region~\cite{sinha2021PRB}. 

This model is prepared in the all-down ground state $\ket{\psi_0}=\ket{\downarrow\downarrow... \downarrow}$, and is driven from one ferromagnetic phase $h_z<0$ to the other ferromagnetic phase $h_z>0$ by linearly ramping the longitudinal field $h_z=vt$. We set $0<h_x \ll 1$ such that the system is far away from the SOQPT point. As the system passes the first-order transition point $h_z=0$, the state $\ket{\psi_0}$ becomes a metastable state and will decay toward the new ground state $\ket{\uparrow\uparrow.. \uparrow}$. The decay dynamics is dominated by successive tunneling to bubble subspaces near specific resonant points, and can be effectively described by the Landau-Zener model. The associated Landau-Zener Hamiltonian is obtained by the Schrieffer-Wolff transformation. The gap at these resonant points is shown to scale exponentially with the number of spins flipped in the excited states that degenerate with the metastable state. This behavior of the gap naturally defines a time-scale separation, and allows us to apply the adiabatic-impulse approximation of KZM: the dynamics is adiabatic when the ramping rate $v$ is much smaller than the gap $\Delta^2(h_z)/\hbar$ at a resonant point $h_z$, and frozen when $v\gg \Delta^2(h_z)/\hbar$. 

The quantum spinodal point can be found using Eqs.~(\ref{eq:metastability})-(\ref{eq:quan_spinodal_condition}). With respect to the state $\ket{\psi_0}$, we take the local operator to be $\mcO=\sigma_j^x$ which flips one spin at site $j$. The spinodal condition of Eq.~(\ref{eq:quan_spinodal_condition}) gives 
\begin{equation}
\langle H(h^{\mathrm{sp}}_z) \rangle_{\sigma_j^x\ket{\psi_0} } - \langle H \rangle_{\ket{\psi_0}} = 2h_{z}^{\mathrm{sp}}-4=0 \implies h_{z}^{\mathrm{sp}}=2.
\end{equation}
At $h_{z}^{\mathrm{sp}}=2$, the state $\ket{\psi_0}$ enters the highest subspace and is degenerate with states like $\ket{\dots\downarrow\uparrow_{j}\downarrow\dots}$, $\ket{\dots \downarrow\uparrow_{j}\downarrow\uparrow_{k}\downarrow\dots}$,..., $\ket{\downarrow\uparrow\downarrow\uparrow\dots.\downarrow\uparrow\downarrow}$. The dimension of this subspace is $\frac{\phi^L-(-\phi)^{-L}}{2\phi-1} =0.447*1.61^{L}$ where $\phi=(1+\sqrt{ 5 }) /2$ is the golden ratio~\cite{turner2018NP,turner2018PRB} and $L$ is the site number. The degeneracy of the state goes to infinity as the system size $L$ goes to infinity. These nearby states form a restricted Hilbert space due to the kinetic constraint that no two adjacent spins can be flipped. This constrained Hilbert space has an emergent $\bbZ_2$ translational symmetry, as the state with the most spin flips is $\ket{\downarrow\uparrow\downarrow\uparrow\dots\downarrow\uparrow\downarrow}$ or $\ket{\uparrow\downarrow\uparrow\downarrow\uparrow\dots\downarrow\uparrow}$. Within this subspace, the effective Hamiltonian governing the dynamics near the quantum spinodal point can be derived using the Schrieffer-Wolff transformation~\cite{bravyi2011AoP}. Up to the leading order of $h_x$, the effective Hamiltonian is the PXP model in a magnetic field~\cite{Vodeb2025NP}
\begin{align}
H_{\mathrm{PXP}} & = -(h_z-h_z^{\mathrm{sp}})\sum_{j}\sigma_{j}^z-h_{x}\sum_{j} P_{j-1}^{\downarrow }\sigma_{j}^x P_{j+1}^\downarrow,
\end{align}
where $P^{\downarrow}=\frac{1-\sigma^z}{2}$ projects the state onto the spin-$\downarrow$ space. 


\begin{figure*}[t]
\includegraphics[scale=1.0]{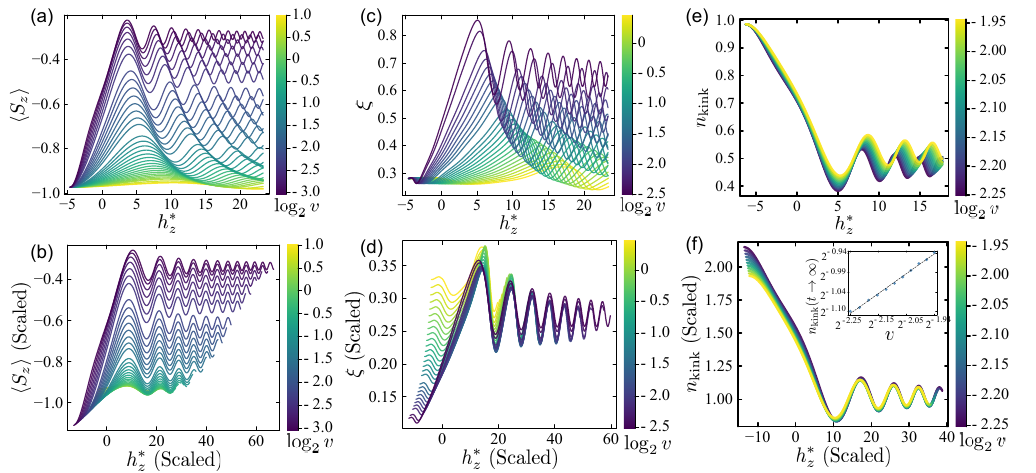}\caption{Time evolution of the tilted Ising model for different ramping rates (encoded by the color scale). Here, $h_z^* = (h_z-h_z^{\mathrm{sp}})/h_x$. Panels (a) and (b) show the magnetization $S_z$ of the tilted Ising model in the unscaled and scaled variables, respectively. Panels (c) and (d) display the unscaled and scaled correlation length $\xi$. Panels (e) and (f) display the unscaled and scaled topological defects. Because $S_z$ is not an order parameter of the effective Hamiltonian, it does not obey a strict scaling ansatz, and only exhibits peaks at the same positions. In contrast, the correlation length and the topological defects collapse in a better way, as shown in panels (c)-(f). The inset in panel (f) shows the KZ scaling of the topological defects at the final time with the fitting to be $n_{\mathrm{kink}}\sim v^{0.52}$, close to the theoretical prediction of $v^{1/2}$ for the Ising universality class. All calculations are performed by the TDVP algorithm with a maximum bond dimension of $200$ and a time step of $0.01$. The system size is $L=259$ and the transverse field is $h_x=0.125$. 
}
\label{fig:titledIsing}
\end{figure*}

The PXP model in a magnetic field is known to exhibit a SOQPT with the Ising universality at $(h_z-h_z^{\mathrm{sp}}) \approx 0.655 h_x$~\cite{sachdev2002PRB,fendley2004PRB,yao2022PRB}. Therefore, the spinodal point is strictly not the transition point. The order parameter is the staggered magnetization $S_s=\frac{1}{L}\sum_{j}\avg{\sigma^z_{j}}(-1)^j$, instead of the order parameter of the FOQPT: magnetization $S_z=\frac{1}{L}\sum_{j}\avg{\sigma^z_{j}}$. The staggered magnetization should follow the finite-time scaling~\cite{gong2010NJP, Zhong2011book, huang2014PRB} 
\begin{align}
    S_{s}(h) = v^{\beta/(r\nu)}f(hv^{-1/(r\nu)}),
\end{align}
where $f(x)$ is a non-universal scaling function and $r=z+1 /\nu$. We substitute the Ising critical exponents
$\nu =1$, $z=1$, $\beta=1/8$ and obtain 
\begin{align}
    S_{s}(h) = v^{1/16}f(hv^{-1/2}).
    \label{eq:scale_tilted}
\end{align}
Unfortunately, the staggered magnetization remains $0$ since the system goes from the disordered phase to the ordered phase. Instead, we use the original order parameter $S_z$ to verify the scaling of the argument $hv^{-1/2}$. The evolution of $S_z$ and scaled $S_z$ are plotted in Fig.~\ref{fig:titledIsing}(a) and (b), respectively. The scaled $S_z$ curves of different $v$ exhibit peaks at the same positions, confirming the predicted scaling relation $h_z$ (see Fig.~\ref{fig:titledIsing}(b)). The scaled value of $S_z$ does not collapse exactly which is unsurprising since $S_z$ is not the true order parameter of the effective Hamiltonian. We also look at the correlation length $\xi$ and the topological defects $n_{\mathrm{kink}}$ which are more directly related to the criticality. The topological defects are defined as $n_{\mathrm{kink}}=\frac{1}{2L}\sum_{j}(1+\langle\sigma^z_{j}\sigma^z_{j+1}\rangle)$. Both of them collapse to a single curve very well after rescaling in Fig.~\ref{fig:titledIsing}(c)-(f). The KZ scaling of the topological defects at the final time is shown in the inset of Fig.~\ref{fig:titledIsing}(f), which gives $n_{\mathrm{kink}}\sim v^{0.52}$, close to the theoretical prediction of $v^{1/2}$ for the Ising universality class. These results confirm that the dynamics near the quantum spinodal point of the 1D tilted Ising model is governed by an emergent SOQPT with Ising universality.

\textit{Example 2: PXP model.}--- 
Our second example is the 1D PXP model under a staggered magnetic field. This example shows that not all FOQPTs feature emergent SOQPTs near their quantum spinodal points but can still be understood by our theory. The Hamiltonian reads
\begin{equation}
H = \lambda \sum_{j} P_{j}^{\downarrow}\sigma^x_{j+1}P_{j+1}^{\downarrow} - \sum_{j}\sigma^z_{j} - h_{s} \sum_{j}(-1)^j \sigma^z_{j}
\end{equation}
with $\lambda \ll 1$. The staggered field $h_s$ breaks its $\mathbb{Z}_{2}$ translational symmetry and induces a FOQPT at $h_s=0$ which is verified in \cite{supp}. The order parameter is the staggered magnetization $S_{s}$. 

We prepare the system in the ground state $\ket{\uparrow\downarrow\uparrow\downarrow\dots}$ at $h_{s}< 0$ and ramp the staggered field $h_s=vt$. We  take the local operator to be $\mcO=\sigma_{j}^x$. First note that the constraint on the Hilbert space also constrains the operator $\sigma_{j}^x$. If we apply the $\sigma_{2j}^x$, then it will create a state which has two adjacent spins pointing up. This state is forbidden by the constraint of the Hilbert space. On the other hand, the spin operator on the odd sites $\{\sigma_{2j+1}^x\}$ can be freely applied to the state. Thus, we only need to consider the operators $\{\sigma_{2j+1}^x\}$ on odd sites. Then the energy change by the action of spin-flip operators on the odd site is 
\begin{equation}
\langle H \rangle_{\mathcal{O}\ket{\psi} } -\langle H \rangle_{\ket{\psi}} = 2-2h_{s} + \mathrm{O}(\lambda).
\end{equation}
It gives the spinodal point at $h_{s}^{\mathrm{sp}}=1$. Since the operators are only allowed to act on the odd sites and thus satisfies the non-adjacent condition automatically, no $\mathbb{Z}_{2}$ translational symmetry is induced, unlike in Example 1. In other words, the Hilbert space constraint prevents the emergence of the translational $\mathbb{Z}_{2}$ symmetry and no effective SOQPT emerges.

Indeed, we expand the Hamiltonian around the spinodal point $h^{\mathrm{sp}}_{s}=1$ and obtain
\begin{equation}
H = \sum_j \lambda\sigma^x_{2j+1} - \delta_s  \sigma^z_{2j+1} + \mathrm{O}(\lambda^2).
\label{eq:ham_pxp_eff}
\end{equation}
where $\delta_s\equiv h_s-h_s^{\mathrm{sp}}$ and the even-site spins are fixed to $\downarrow$. The odd-site spins are completely decoupled and there is no phase transition. The dimension of the reduced Hilbert space is $2^{L/2}$. 

\begin{figure}
\includegraphics[scale=1.0]{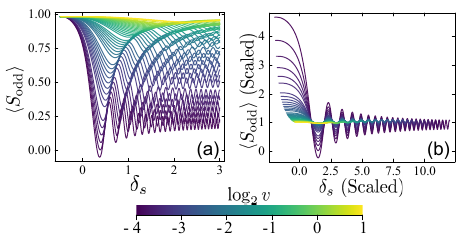}\caption{Magnetization at odd sites for PXP model with a staggered field. (a) Evolution for the original Hamiltonian, calculated by TDVP
method. We consider a system with 259 sites, starting from $\delta_s = -0.5$ at its metastable state. Another parameter is $\lambda = 0.1$. (b) Scaled evolution. We scale the vertical axis as $S_\mathrm{odd}/[2\exp(-\pi \lambda^2/v)-1]$ and find the final value of all curves asymptotically approach 1, consistent with the Landau-Zener theory. We
scale the horizontal axis as $v^ {0.5}\delta_s $.
\label{fig:pxp_model}}
\end{figure}

The Hamiltonian~(\ref{eq:ham_pxp_eff}) is exactly solvable. The time-dependent Schrödinger equation can be decomposed into $L/2$ independent Landau-Zener problems. The transition probability of each spin is given by the Landau-Zener probability $P_{LZ} = \exp(-\pi \lambda^2/v)$. A relevant physical quantity is the magnetization at odd sites $S_\mathrm{odd}=(L/2)^{-1}\sum_{j}\langle\sigma^z_{2j+1}\rangle$. The predicted final magnetization is
\begin{equation}
S_{\mathrm{odd}} = 2P_{LZ}-1 = 2\exp(-\pi \lambda^2/v) - 1.
\end{equation}
This is verified numerically in Fig.~\ref{fig:pxp_model} that the scaled magnetization $S_{\mathrm{odd}}(\mathrm{scaled})$ converges to the same line.

\emph{Summary and outlook.}---We propose a microscopic framework for understanding the emergence of criticality in the vicinity of the spinodal point of FOQPTs. 
An important notion of our work is that the breakdown of $(\Delta,R)$-metastability is governed by a specific class of local operators. These operators naturally enforce kinetic constraints on the high-energy resonant subspace. As demonstrated by our first paradigmatic example, the universality class of the effective SOQPT is determined entirely by the symmetries of this effective Hamiltonian. In the 1D tilted Ising model, the constraints map the spinodal dynamics to the PXP model, leading to an emergent translational $\bbZ_2$ symmetry breaking and Ising universality. In our second example the PXP model with a staggered field, the intrinsic Hilbert space constraints prohibit the emergence of the translational symmetry, leading to the absence of both an effective SOQPT and the criticality. 

Our theory provides a universal mechanism for understanding the dynamics of FOQPTs beyond the mean-field theory. Recent advances in cold-atom platforms, particularly Rydberg atom arrays, allow the blockade radius to be tuned to engineer specific symmetries~(e.g., $\bbZ_n$). These systems thus provide a natural and controllable setting to test our predictions~\cite{Langen2015arcm,henriet2020quantum}. This highlights that current programmable quantum simulators are ideally suited to experimentally verify these “hidden” phase transitions. Future work could extend our analysis to systems with long-range interactions. In the infinite-range limit, interactions drive the system toward mean-field behavior, characterized by a power-law scaling of $v^{2/5}$ \cite{Bapst2012JSP}. An interesting open question concerns high-dimensional systems with local interactions that nevertheless admit a mean-field description, such as the three-dimensional tilted quantum Ising model. 
More generally, it would be valuable to understand how long-range interactions modify the universality class near the quantum spinodal point.


\emph{Acknowledgments}.---We thank Yu-Xing Wu and Fan Zhong for helpful discussions. Chiao Wang and H. T. Quan acknowledge the support from the National Science Foundation of China under grants 12375028 and 12521004.

\bibliographystyle{apsrev4-2}
\bibliography{ref}

\end{document}


\preprint{APS/123-QED}

\title{\textbf{Supplementary Material: Criticality at the Spinodal Point of First-Order Quantum Phase Transitions} 
}%
\author{Fan Zhang}
\email{van314159@pku.edu.cn}
\affiliation{School of Physics, Peking University, Beijing, 100871, China}

\author{Chiao Wang}
\affiliation{School of Physics, Peking University, Beijing, 100871, China}

\author{H. T. Quan}
\email{htquan@pku.edu.cn}
\affiliation{School of Physics, Peking University, Beijing, 100871, China}
\affiliation{Collaborative Innovation Center of Quantum Matter, Beijing 100871,
China}
\affiliation{Frontiers Science Center for Nano-optoelectronics, Peking University,
Beijing, 100871, China}

\date{\today}

\begin{abstract}
In this Supplementary Material, we provide detailed theoretical derivations and numerical evidence to support the findings in the main text.
In Sec.~\ref{sec:FTS}, we first review the Kibble-Zurek mechanism and then introduce the finite-time scaling (FTS) framework, demonstrating its consistency with the Kibble-Zurek mechanism.
In Sec.~\ref{sec:case study}, we apply FTS to both the 1D transverse-field Ising model (TIM) and the PXP model. We analytically derive the scaling relations for the TIM, and numerically verify the FTS predictions for both models using TDVP simulations.
In Sec.~\ref{sec:pxp_model_prop}, we characterize the first-order phase transition in the PXP model, where we identify the spinodal point and derive an effective Landau-Zener description for the dynamics using the Schrieffer-Wolff transformation.
Finally, in Sec.~\ref{sec:ppxpp_model_prop}, we extend our analysis to the one-dimensional next-nearest neighbor (NNN) tilted Ising model. We demonstrate that its dynamics near the quantum spinodal point are governed by an emergent effective PPXPP model, and we verify the validity of FTS for this emergent $\mathbb{Z}_3$ symmetry-breaking phase transition.
\end{abstract}

\maketitle

\tableofcontents

\section{\label{sec:FTS}
Kibble-Zurek Scaling and Finite-time Scaling}

In this section, we review the Kibble-Zurek mechanism (KZM) and the finite-time scaling (FTS) ansatz for second-order phase transitions. The KZM describes the universal non-equilibrium dynamics of a system driven across a second-order phase transition~\cite{Kibble1976jpa, Zurek1985nature, Zurek2005prl,Uhlmann2007prl}. At its core, KZM relies on the adiabatic-impulse-adiabatic approximation. Far from the critical point, the energy gap $\Delta$ is large, and the relaxation time $\tau_r \sim \Delta^{-1}$ is short; consequently, the system evolves adiabatically by following the instantaneous Hamiltonian. In this adiabatic regime, the system remains in its instantaneous ground state. However, as the system approaches the critical point, the gap vanishes as $\Delta \sim |\lambda|^{z\nu}$, causing the relaxation time and the correlation length to diverge as $\tau_r \sim |\lambda|^{-\nu z}$ and $\xi \sim |\lambda|^{-\nu}$, respectively. Here, $\lambda$ is a dimensionless parameter characterizing the deviation from the critical point~(CP), while $\nu$ and $z$ are the static and dynamic critical exponents~\cite{sachdev2011Book}. Due to this critical slowing down, there inevitably exists a moment when the relaxation time $\tau_r$ becomes comparable to the time scale of the quench, $|\lambda/\dot{\lambda}|$. At this point, the system fails to keep pace with the change of the Hamiltonian. KZM approximates this breakdown of adiabaticity by introducing an impulse regime, assuming the state effectively freezes until the system exits the impulse region. The boundary of this frozen regime is determined by the condition where the relaxation time matches the quench timescale, i.e., $\tau_r(\lambda^*) \sim |\lambda^*/\dot{\lambda}|$. For a linear quench where $\lambda(t) = vt$ (with $t$ running from $-\infty$ to $\infty$), the freeze-out parameter $\lambda^*$ scales as~\cite{Dziarmaga2010AP}:
\begin{equation}
\lambda^* \sim v^{\frac{1}{\nu z+1}}.
\end{equation}
Correspondingly, the freeze-out time scales as $t^{*} \sim v^{-\nu z/(\nu z+1)}$, and the characteristic length scale as $\xi^{*} \sim v^{-\nu/(\nu z+1)}$. Finally, the density of topological defects $n$ is estimated by:
\begin{equation}
n \sim \xi^{-d} \sim v^{\frac{d\nu}{\nu z+1}}. 
\label{eq: n_KZM}
\end{equation}


While the KZM focuses on the scaling relation at the final time, the FTS ansatz extends this idea to a finite-time evolution process~\cite{gong2010NJP}. Notably, the scaling relations derived from KZM can be recovered as a special case within this FTS framework. Let us consider a classical system characterized by an order parameter $M$, universality dictates the following scaling ansatz in the vicinity of the critical point: 
\begin{equation}
M(t,\tau,h) = b^{-\beta/\nu} f(tb^{-z}, \tau b^{1/\nu}, hb^{\beta\delta/\nu}),
\end{equation}
where $\tau$ denotes the reduced temperature, and $h$ is the external field deviation from the critical point (we consider the linear driving protocol $h = vt$ for example). The validity of the FTS ansatz requires a specific hierarchy of timescales. The characteristic timescale of the driving process must significantly exceed the intrinsic microscopic evolution timescale of the system ($\tau_{\text{drive}} \gg \tau_{\text{sys}}$).

It's instructive to compare the FTS with the well-known finite-size scaling (FSS). FSS arises when the divergence of the correlation length $\xi$ is truncated by the finite system size $L$, rendering $L$ a relevant scaling variable. Analogously, FTS originates from the divergence of the relaxation time $\tau_r$ near the critical point. In a dynamic process, the external driving imposes a characteristic timescale $\tau_{\text{drive}}$. This driving timescale effectively cuts off the divergence of $\tau_r$, playing a role similar to the system size $L$ in FSS.

The FTS theory has already been rigorously verified in classical phase transitions~\cite{zhong2005PRL,Huang2014PRB}. This framework naturally extends to quantum phase transitions (QPTs)~\cite{Yin2014prb}. Unlike classical transitions driven by thermal fluctuations (temperature $\tau$), QPTs occur at zero temperature and are driven by quantum fluctuations associated with a non-thermal control parameter $g$ in the Hamiltonian. For instance, in the 1D transverse-field Ising model in Sec.~\ref{subsec:TIM_universality}, $g$ represents the normalized deviation of the transverse field from its critical value, i.e., $g = (h_x/J) - (h_x/J)_c$ where $h_x/J=1$ is the critical point of the 1D transverse-field Ising model. Consequently, the FTS ansatz for a quantum system is obtained by substituting the thermal variable $\tau$ with the quantum control parameter $g$:
\begin{equation} \label{eq: FTS scaling ansatz quantum}
M(t,g,h_z) = b^{-\beta/\nu} f(t b^{-z}, g b^{1/\nu}, h_z b^{\beta\delta/\nu}).
\end{equation}
where we have identified the symmetry-breaking field with the longitudinal field $h_z$. This ansatz assumes that the characteristic timescale of the driving process significantly exceeds the intrinsic microscopic timescale of the system ($\tau_\text{drive} \gg \tau_\text{sys}$), ensuring that time $t$ emerges as a relevant variable in the renormalization group framework. Now we apply this FTS ansatz to analyze the density of topological defects in the quantum context. From dimensional analysis, the density of defects has the dimension $[L]^{-d}$. We expect it to transform as $n'(t', g') = b^{d}n(t, g)$ under a scale transformation. This yields the FTS ansatz for $n$:
\begin{equation}
n(t,g) = b^{-d} f(g b^{1/\nu}, tb^{-z}),
\end{equation}
where we have replaced $\tau$ with the parameter $g$ as discussed, and set the symmetry-breaking field $h_z = 0$. Considering a linear protocol $g=v t$, the rescaling of the variables follows:
\begin{equation}
g' = v't' = g b^{1/\nu} \implies v' = vb^{(z\nu+1)/\nu} = vb^{r},
\end{equation}
where $r \equiv z+1 /\nu$. By choosing the scaling factor $b=v^{-1/r}$, we obtain:
\begin{equation}
n(g) = v^{d/r}\tilde{f}(g v^{-1/(r\nu)}) = v^{d\nu/(z\nu+1)} \tilde{f}(g v^{-1/(z\nu+1)}).
\end{equation}
As $g \to \infty$, the scaling function approaches a constant, leading to:
\begin{equation}
n \propto v^{d\nu/(z\nu+1)},
\end{equation}
which is consistent with the result obtained by the KZ scaling in Eq.~(\ref{eq: n_KZM}).

The scaling of the topological defect density derived above is fundamentally rooted in the divergence of the correlation length $\xi$. Recalling the core Kibble-Zurek relation $n \sim \xi^{-d}$, this connection provides a natural pathway to establish explicit universal scaling relations for other thermodynamic quantities under finite-time driving. For the correlation length, the standard scaling ansatz dictates $\xi(t,g) = b f(g b^{1/\nu}, t b^{-z})$. By substituting the linear quench protocol $g=vt$ and choosing the scaling factor $b = v^{-1/r}$, the explicit time dependence is absorbed. This yields the universal scaling relation for the correlation length:
\begin{equation}
\xi v^{\frac{1}{r}} = f\left(g v^{-\frac{1}{\nu r}}\right).
\end{equation}
We can also explicitly derive the dynamic scaling law for the order parameter $M$. Starting from the general quantum scaling ansatz introduced earlier and setting the conjugate symmetry-breaking field to zero, the relation Eq.~(\ref{eq: FTS scaling ansatz quantum}) reduces to $M(t,g) = b^{-\beta/\nu} \tilde{f}(g b^{1/\nu}, t b^{-z})$. Employing the exact same linear driving relation $g=vt$ and scaling factor $b = v^{-1/r}$, we arrive at the definitive FTS form for the order parameter:
\begin{equation}
M v^{-\frac{\beta}{\nu r}} = f\left(gv^{-\frac{1}{\nu r}}\right).
\end{equation}
These fundamental results demonstrate that both the correlation length and the magnetization trajectories obtained under different quench velocities will collapse onto their respective universal curves when the physical observables and the control parameter are rescaled by the appropriate theoretical powers of $v$.

In summary, the FTS framework establishes a highly self-consistent theoretical description that intrinsically encompasses the KZM. Deeply rooted in the underlying criticality of the system, FTS goes beyond merely predicting asymptotic defect densities by providing a comprehensive description of the universal scaling behavior throughout the entire non-equilibrium evolution process. Consequently, it provides a robust and powerful tool to investigate the complex dynamics of quantum phase transitions.


\section{\label{sec:case study}
Model-specific analysis of finite-time scaling}

In this section, we provide a detailed model-specific analysis to validate the FTS ansatz in quantum many-body systems. First, in Sec.~\ref{subsec:TIM_universality}, we introduce the one-dimensional transverse-field Ising model (TIM) and derive the expected scaling relations based on its critical exponents. Then, in Sec.~\ref{subsec:TDVP_sim}, we present numerical results obtained via Time-Dependent Variational Principle (TDVP) simulations for both the TIM and the PXP model, which show excellent agreement with our theoretical predictions.

\subsection{\label{subsec:TIM_universality}FTS from dimensional analysis: 1D TIM}

In this subsection, we derive the specific finite-time scaling relations for the 1D TIM. As a prototypical model for quantum phase transitions, the TIM allows for an exact analytical treatment via the Jordan-Wigner transformation, which maps the spin chain to a system of free fermions. By identifying the static and dynamic critical exponents characteristic of the 1D Ising universality class, we apply the general FTS ansatz to explicitly construct the scaling forms for both the longitudinal magnetization $m_z$ (the order parameter) and the correlation length $\xi$.

The Hamiltonian of the 1D TIM is  
\begin{align}
	H=-J\sum_{j}\sigma_{j}^{z}\sigma_{j+1}^{z}-h_{x}\sum_{j}\sigma_{j}^{x}, 
\end{align}
where $\sigma_{j}^{x,y,z}$ is the Pauli spin-$1 /2$ matrix at site $j$, $J>0$ is the ferromagnetic interaction, and $h_{x}$ is the transverse field. This model has an order-disorder quantum second-order phase transition at $|h_x/J|=1$ which separates the ferromagnetic phase $|h_x/J|<1$ and paramagnetic phase $|h_{x}/J|>1$.

The finite-time scaling relation can be derived solely from the dimensional analysis and the critical exponents. We denote the correlation length as $\xi$, $g$ as the reduced temperature or the relative distance away from the critical point, and denote $m$ and $h$ as the order parameter and the symmetry-breaking field. In TIM, $g,m,h$ are given by
\begin{equation}
g=(h_x /J)- (h_x /J)_{c},\quad m=m_{z},\quad h=h_{z},
\end{equation}
where $(h_x /J)_{c}=1$ is the critical point and $h_z$ is the longitudinal field added here for completeness. The idea is to invoke the scale ansatz around the critical point. It means if we perform a scale transformation $\xi'=\xi b^{-1}$ where $b$ is the scaling factor, then other physical quantities should transform as
\begin{equation}
g'=(\xi')^{-1/\nu}=gb^{1/\nu},\quad h_{z}'=h_{z}b^{\beta \delta/\nu},\quad t'=tb^{-z},\quad m_{z}'=m_{z}b^{\beta/\nu}, 
\end{equation}
where $\nu,$ $\beta,$ $\delta,$ are the static critical exponents and $z$ is the dynamic critical exponent. If we write $m_z$ as the function of $g,h_z,t$, then the above scaling relation implies
\begin{equation}
m_{z}(g,h_{z},t) = b^{-\beta/\nu}m_{z}'(g',h_{z}',t')=b^{-\beta /\nu}m_{z}'(gb^{1/\nu},h_{z}b^{\beta\delta/\nu},tb^{z}).
\end{equation}
Since we are focusing on the TIM, we set $h_z=0$ and adopt the linear protocol $g=vt$ which is commonly used in the KZM. Then from $g'=b^{1/\nu}g$, we can determine the transform rule of $v$ as
\begin{equation}
v't' = b^{1/\nu}vt \implies v'=b^{1/\nu+z}v.
\end{equation}
Denote $r=1 /\nu+z$, take $b=v^{-1/r}$ and fix $v'=1$. Then we obtain the finite-time scaling relation for $m_z$:
\begin{equation}
m_{z}(g) = v^{\beta/(\nu r)}f(gv^{-1/(\nu r)}) = v^{\beta/(\nu r)}f(gv^{-1/(1+z\nu)}).
\end{equation}
Substituting the critical exponents of the 1D TIM 
\begin{equation}
\nu =1, \quad z=1, \quad \beta=\frac{1}{8},\quad \gamma = \frac{7}{4},\quad \delta=15,
\end{equation}
we find
\begin{equation}
m_{z}(g) = v^{1/16}f_{z}(gv^{-1/2}).
\end{equation}

We can also consider $\xi$ which is equivalent to the  density of topological defects $n$ which scales as $n \sim \xi^{-d}$ with $d=1$ being the spatial dimension in the 1D TIM. Then we have the finite-time scaling relation for $\xi$:
\begin{align}
\xi(g,h_{z},t) &= b^{-d}\xi'(g',h_{z}',t')=v^{d/r}f_{\xi}(gv^{-1/(1+z\nu)}) = v^{d\nu/(1+z\nu)}f_{\xi}(gv^{-1/(1+z\nu)})\\
 & = v^{1/2}f_{\xi}(gv^{-1/2}) \label{Eq:fts_xi_TIM}
\end{align}
If we only consider the final time $t\to \infty$, then $g\to \infty$ and the scaling function approaches a constant, we recover the KZ scaling for the density of topological defects in the 1D TIM~\cite{dziarmaga2005PRL}.

\subsection{\label{subsec:TDVP_sim}
TDVP Simulation: 1D TIM and 1D PXP model}

In this subsection, we present the numerical verification of our FTS theory using the Time-Dependent Variational Principle (TDVP) based on Matrix Product States (MPS)~\cite{fishman2022, ITensorBase}. We focus on two distinct models: the standard 1D TIM and the 1D PXP model (with a longitudinal field).

While the critical behavior of the TIM is well-established, the PXP model requires a brief introduction. Originating from the description of Rydberg atom arrays in the blockade regime, the PXP model is governed by the Hamiltonian:
\begin{equation}
\label{eq:ham_pxp_std}
H_{PXP} = \sum_{j} P_{j-1}^\downarrow \sigma^x_j P_{j+1}^\downarrow - h_z \sum_{j} \sigma^z_j,
\end{equation}
where $P_j^\downarrow = (1-\sigma^z_j)/2$ is the projector onto the ground state $\ket{\downarrow}_i$, enforcing the constraint that no two adjacent atoms can be simultaneously in the excited state $\ket{\uparrow}$. Although the PXP model is famous for its weak ergodicity breaking (quantum scars) at $h_z=0$~\cite{bernien2017Nature, turner2018NP}, it hosts a quantum phase transition when the detuning field $h_z$ is varied. Specifically, at the critical point $(h_z)_c \approx 0.655$, the system undergoes a second-order phase transition from a disordered phase to a $\mathbb{Z}_2$ symmetry-breaking antiferromagnetic phase. Crucially, despite the kinetic constraint, this transition falls into the (1+1)D Ising universality class, sharing the same critical exponents ($\nu=1, z=1, \beta=1/8$) as the TIM~\cite{Yao2022PRB}.

We perform linear quench simulations $g(t) = vt$ for both models starting from the ground state. For the TIM, the control parameter is $g = h_x(t) - 1$ (with $J=1$), and the order parameter is the longitudinal magnetization $S_z = \sum_j \langle \sigma^z_j \rangle / N$. For the PXP model, the control parameter is defined as the deviation from its critical point $g = h_z(t) - 0.655$, and the appropriate order parameter is the staggered magnetization $S_{sta} = \sum_j (-1)^j \langle \sigma^z_j \rangle / N$.

\begin{figure}
	\includegraphics[scale=0.5]{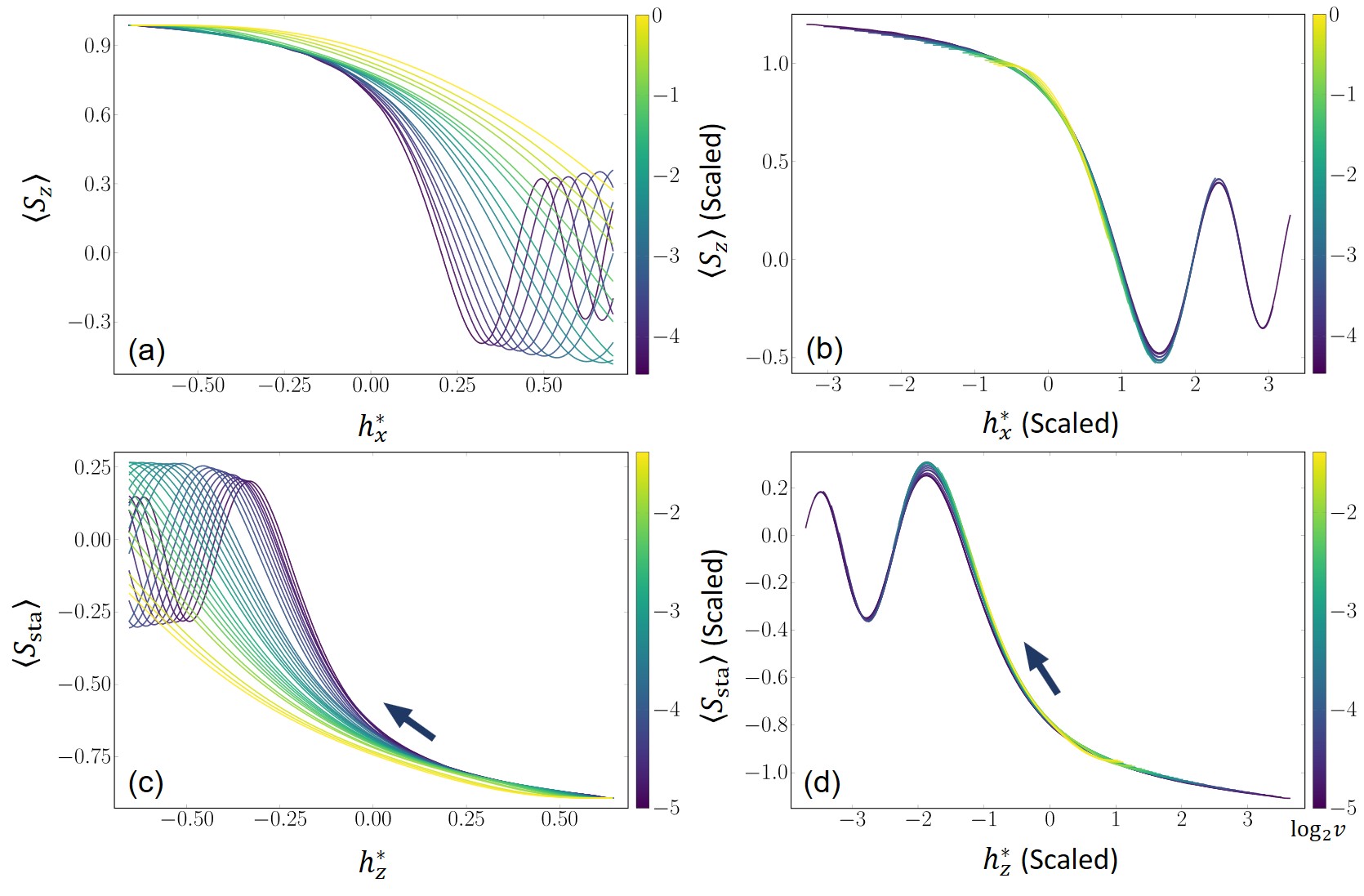}
	\caption{Finite-time scaling of order parameters in the 1D TIM (top row) and PXP model (bottom row), calculated via the TDVP method with open boundary conditions (OBC) and $N=97$ spins. (a) Original dynamics of the longitudinal magnetization $S_z$. (b) Data collapse of the curves in (a) according to the FTS ansatz $S_z=v^{1/16}f(gv^{-1/2})$, where the reduced control parameter is $g \equiv h_x^* = h_x - 1$. (c) Original dynamics of the staggered magnetization $S_\text{sta} = \sum_j (-1)^j \sigma^z_j /N$ in the PXP model near its second-order phase transition. (d) Corresponding data collapse using the FTS ansatz with the same Ising critical exponents, with $g \equiv h_z^* = h_z - 0.655$.
	\label{fig:supp_TIM}}
\end{figure}

The numerical results are summarized in Fig.~\ref{fig:supp_TIM}. We examine the dynamic scaling behavior of the order parameter for the 1D TIM (top row) and the PXP model (bottom row). Based on the critical exponents of the 1D Ising universality class ($\nu=1, z=1, \beta=1/8$), the finite-time scaling ansatz predicts that the magnetization curves should follow the scaling form $M(t) = v^{1/16} f(g v^{-1/2})$. Figures~\ref{fig:supp_TIM}(a) and (c) display the original time-evolution trajectories under various quench velocities. To validate the theoretical prediction, we perform a data collapse analysis by rescaling the axes according to the FTS ansatz. As shown in Figs.~\ref{fig:supp_TIM}(b) and (d), by plotting the rescaled magnetization $ S_z v^{-1/16}$ against the rescaled distance to criticality $g v^{-1/2}$, the curves for different velocities collapse onto a single curve. This perfect collapse confirms the validity of the FTS framework.

\begin{figure}
	\includegraphics[scale=0.5]{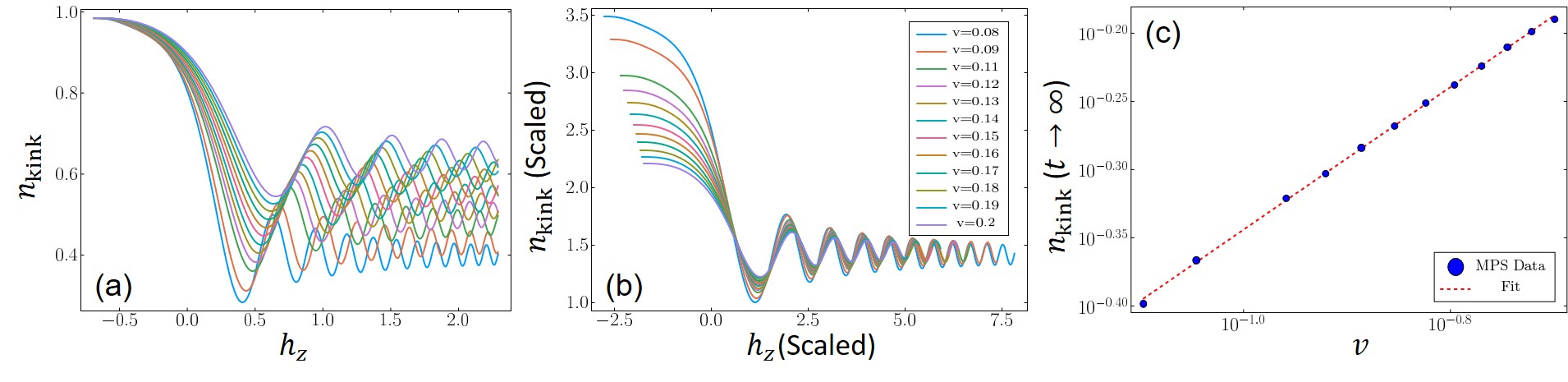}
	\caption{Kibble-Zurek scaling of the kink density in the PXP model across its second-order phase transition. The dynamics are calculated using the TDVP method on a chain of $N=259$ sites, with the detuning field linearly quenched from an initial value of $h_{zi}=-0.7$. (a) Original time evolution of the unscaled kink density $n_{kink} = \frac{1}{N} \sum_i (1 + \sigma_i^z \sigma_{i+1}^z) / 2$ as a function of the driving parameter $h_z$ for various quench velocities $v$. (b) Data collapse of the trajectories from (a) according to the finite-time scaling ansatz, plotting the scaled kink density against the scaled distance to criticality $(h_z - h_c)v^{-0.5}$. (c) Log-log plot of the asymptotic kink density $n_{kink}(t \to \infty)$ versus the quench velocity $v$. The red dashed line represents a linear fit to the data, yielding a scaling exponent of $\mu \approx 0.52$. This result is in excellent agreement with the theoretical KZM prediction of $\mu = 0.5$ for the $(1+1)$D Ising universality class.
	\label{fig: kzm scaling in pxp sopt}}
\end{figure}

Furthermore, we investigate the Kibble-Zurek mechanism (KZM) by calculating the topological defect density generated during the second-order phase transition of the PXP model, as a comparison. In the context of the $\mathbb{Z}_{2}$ symmetry-breaking antiferromagnetic phase, we define the kink density operator as
\begin{equation}
n_{kink} = \frac{1}{N} \sum_j (1 + \sigma_j^z \sigma_{j+1}^z) / 2.
\end{equation}
According to the KZM scaling relation for the $(1+1)$D Ising universality class, the asymptotic defect density is predicted to scale with the quench velocity $v$ as $n_{kink} \sim v^\mu$ where $\mu=\nu/(1+\nu z)$. Given the critical exponents $\nu=1$ and $z=1$ , the theoretical expectation for the exponent is $\mu = 0.5$. We perform the time evolution using the TDVP method on a chain of 259 sites with $$H= \frac{1}{8}\sum_{j} P_{j-1}^\downarrow \sigma^x_j P_{j+1}^\downarrow - h_z \sum_{j} \sigma^z_j,$$ linearly quenching the detuning field from an initial value $h_{zi} = -0.7$. Here we intentionally set the coefficient of the $\sigma_j^x$ term to be $1/8$ to match the effective Hamiltonian in the main text. The results are summarized in Fig.~\ref{fig: kzm scaling in pxp sopt}. Panel (a) displays the original time evolution of the kink density for various quench velocities. Panel (b) illustrates the scaled version of these trajectories, demonstrating a clear data collapse that further validates the FTS ansatz. Finally, panel (c) shows a linear fit to the asymptotic kink density against the quench velocity on a logarithmic scale; the numerical fitting yields an exponent of $\mu = 0.52$, which is in agreement with the theoretical prediction of 0.5.

\section{\label{sec:pxp_model_prop}
First-order phase transition in the 1D PXP model in a staggered field}

In this section, we introduce an extended 1D PXP model that exhibits a clear quantum first-order phase transition. We provide a detailed analysis of its static properties, the location of the phase transition and spinodal points, and the effective description derived via the Schrieffer-Wolff transformation (SWT).

\subsection{Hamiltonian and Evidence of First-order Phase Transition}

The standard PXP model is known to host a continuous Ising-type phase transition driven by the detuning field $h_z$. In the ordered phase ($h_z > h_c$), the ground state is doubly degenerate, corresponding to the two possible $\mathbb{Z}_2$ antiferromagnetic configurations, $\ket{\mathbb{Z}_2} = \ket{\uparrow\downarrow\uparrow\downarrow\dots}$ and $\ket{\mathbb{Z}_2'} = \ket{\downarrow\uparrow\downarrow\uparrow\dots}$ (in the Rydberg occupation basis). To induce the first-order phase transition, we introduce a staggered magnetic field $h_s$ which lifts this degeneracy. The Hamiltonian for this extended PXP model (with longitudinal and staggered fields) is defined as:
\begin{equation}
\label{eq:ham_pxp_fopt}
H = J \sum_j P_{j-1}^\downarrow \sigma_{j}^x P_{j+1}^\downarrow - h_z \sum_j \sigma_{j}^z - h_s \sum_j (-1)^j \sigma^z_j.
\end{equation}
We fix $J$ and $h_z$ such that the system is deep within the ordered phase of the standard PXP model (e.g., $h_z > h_c$ in the language of detuning, or specifically $J = 0.1h_z$ as used in our theoretical analysis).

\begin{figure}
	\includegraphics[scale=0.6]{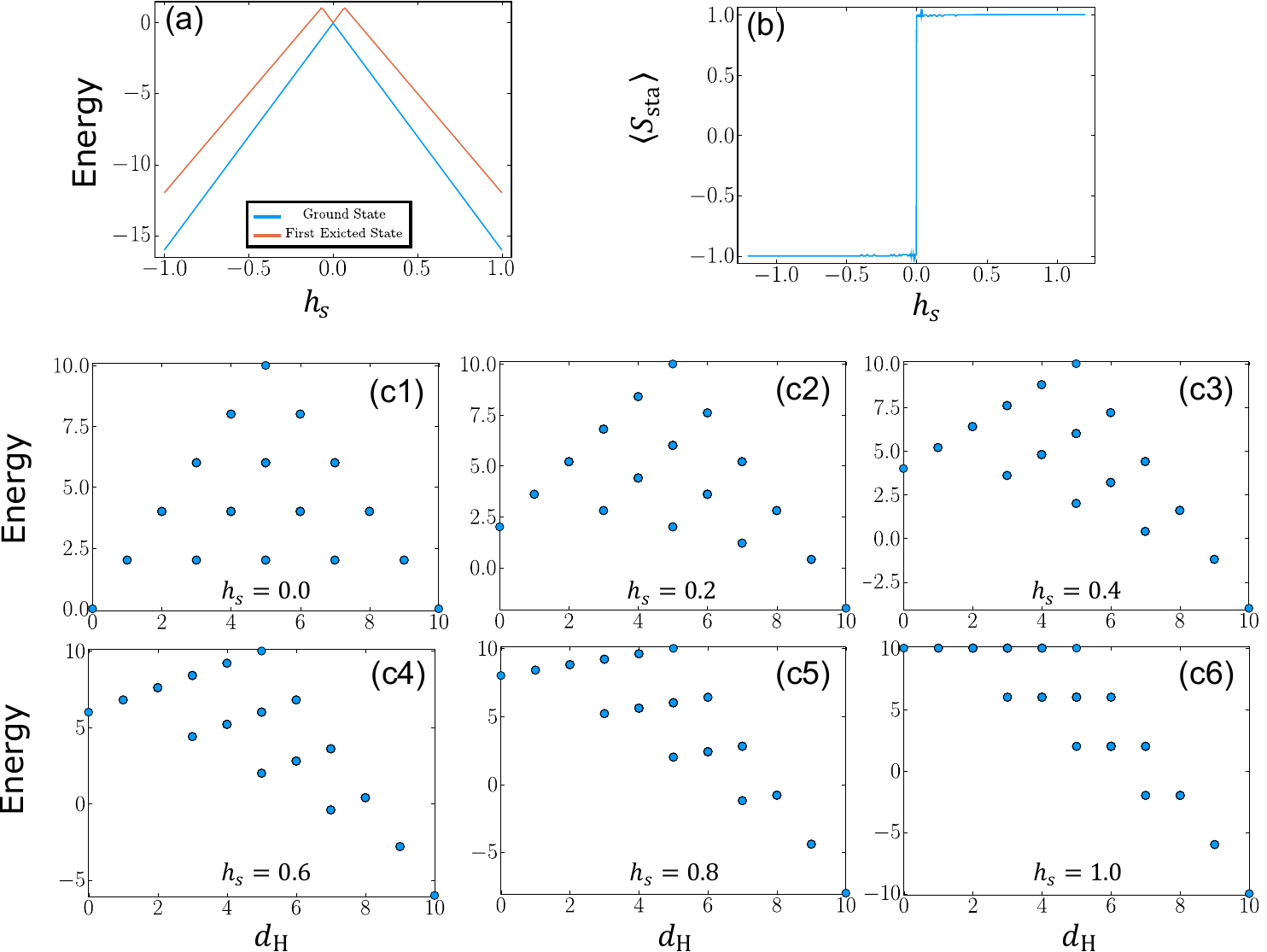}
	\caption{The static properties of the PXP model Eq.~(\ref{eq:ham_pxp_fopt}). (a) The energy of the ground state and the first excited state near the first-order phase transition point $h_{s}=0$. There is a sharp crossover between two states. (b) The staggered magnetization $M_{s}$ also shows a sharp jump at the transition point. (a) and (b) are calculated by the density matrix renormalization group (DMRG). We set $J = 0.1, h_z = 1$ and the site number is $350$. (c) The energy spectrum as a function of the Hamming distance $d_\mathrm{H}$ to $\ket{\mathbb{Z}_2}$ state for different $h_{s}$. Here the site number is $10$. The energy spectrum shows a transition from a double well to a single well.}
	\label{fig:supp_pxp}
\end{figure}

The phase transition driven by sweeping $h_s$ across zero is strictly first-order. This classification is supported by three distinct pieces of evidence:

\begin{enumerate}
    \item \textbf{Direct Level Crossing:} Exact diagonalization results, presented in Fig.~\ref{fig:supp_pxp}(a), reveal a direct level crossing between the ground state and the first excited state at the transition point $h_s=0$. Unlike the second-order transitions which feature an avoided level crossing, the gap here closes exactly, reflecting the distinct symmetry sectors of the competing ground states.
    \item \textbf{Discontinuous Order Parameter:} As shown in Fig.~\ref{fig:supp_pxp}(b), the staggered magnetization $M_s = \frac{1}{N}\sum_i (-1)^i \langle \sigma^z_i \rangle$ exhibits a sharp, discontinuous jump at $h_s = 0$. In the thermodynamic limit, this discontinuity signifies a sudden switch between the $\ket{\mathbb{Z}_2}$ and $\ket{\mathbb{Z}_2'}$ ground states, a hallmark of the first-order quantum phase transition.
    \item \textbf{Metastability:} The energy spectrum analysis reveals the existence of metastable states, as illustrated in Fig.~\ref{fig:supp_pxp}(c). As the system is driven away from $h_s=0$, the ``false" ground state remains a local minimum of the energy landscape up to a critical field strength $h_s^{\mathrm{sp}}$. This results in a ``double-well" to ``single-well" transition in the energy landscape, fundamentally different from the ``single-well" flattening observed in continuous transitions.
\end{enumerate}

\subsection{Phase Transition Point and Spinodal Point}

The phase transition point and the spinodal point are two fundamental features characterizing the first-order quantum phase transition. Here, the phase transition point is located at $h_s = 0$, which is determined by the symmetry of the Hamiltonian Eq.~(\ref{eq:ham_pxp_fopt}). At $h_s=0$, the Hamiltonian is invariant under the translational transformation on the lattice, rendering the two symmetry-breaking ground states $|\mathbb{Z}_2\rangle$ and $|\mathbb{Z}_2'\rangle$ energetically degenerate.

The spinodal point $h_s^{\mathrm{sp}}$ of a metastable state can be found by the method in the main text. Let us take the metastable state $\ket{\mathbb{Z}_2}$ for example. The spinodal point, which marks the breakdown of the metastability, is located at $h_s^{\mathrm{sp}} = h_z$.
To calculate this, we consider the stability of the metastable state against single-spin flips.
Let us examine the energy cost of flipping a single spin on the odd sites in the metastable state.
Deep in the ordered phase ($h_z \gg J$), the dominant energy contribution comes from the field terms $H_\text{field} = -h_z \sigma^z_j - h_s (-1)^j \sigma^z_j$.
For an odd site $j$, the local effective field is $-(h_z - h_s)\sigma^z_j$. The energy gap $\Delta E$ required to flip this spin is proportional to the effective field strength: $\Delta E \propto 2(h_z - h_s)$. The spinodal point is defined where this excitation gap vanishes, i.e., $\Delta E = 0$. This yields the condition:
\begin{equation}
    h_z - h_s^{\mathrm{sp}} = 0 \implies h_s^{\mathrm{sp}} = h_z.
\end{equation}
At this point, the local potential barrier confining the metastable state disappears, and the system becomes unstable against local fluctuations.

\subsection{Effective Landau-Zener Description via SWT}

To quantitatively describe the dynamics near the spinodal point, we employ the Schrieffer-Wolff transformation (SWT) to derive an effective Hamiltonian.

The Schrieffer-Wolff transformation is a standard perturbative technique used in many-body physics to derive an effective Hamiltonian for a separated Hilbert subspace $\mathcal{P}_0$ ~\cite{bravyi2011AoP}. This method is particularly applicable when the system possesses a large energy gap $\Delta$ determined by the unperturbed Hamiltonian $H_0$, and the coupling $V$ between the subspaces is weak ($|V| \ll \Delta$). In our system, the strong longitudinal field creates such a gap, while the transverse hopping term acts as the perturbation. The goal of SWT is to construct a unitary transformation $U = e^S$ (where $S$ is an anti-Hermitian generator) that rotates the Hamiltonian $\tilde{H} = e^{-S} H e^S$ into a block-diagonal form, thereby eliminating the off-diagonal coupling elements up to a desired order in perturbation theory. By choosing the generator $S$ to satisfy the condition $[H_0, S] = -V_\text{od}$ (where $V_\text{od}$ is the off-diagonal part of $V$), one can systematically eliminate virtual transitions to the high-energy manifold. The resulting effective Hamiltonian projected onto the subspace $\mathcal{P}_0$ up to the second order is given by:
\begin{equation}
H_\text{eff} = P_0 H_0 P_0 + P_0 V_\text{d} P_0 + \frac{1}{2} P_0 [S_1, V_\text{od}] P_0,
\end{equation}
where $P_0$ is the projector onto $\mathcal{P}_0$, $V_\text{d}$ is the block-diagonal part of the perturbation (acting within $\mathcal{P}_0$), and $S_1$ is the first-order generator determined by the off-diagonal coupling $V_\text{od}$.

\textbf{Choice of Subspace:}
We define our high-energy subspace $\mathcal{Q}_0$ as the manifold where all spins on the even sites are fixed in the spin-down state $\ket{\downarrow}$, i.e., $P_j^\downarrow \ket{\psi} = \ket{\psi}$ for all even $j$.
This choice is justified by the strong longitudinal field $h_z$. In the vicinity of the spinodal point ($h_s \approx h_z$), the effective field on even sites is $-(h_z + h_s) \approx -2h_z$, which creates a large energy penalty for any excitation (spin-up) on the even sublattice. Conversely, the effective field on odd sites is $-(h_z - h_s) \approx 0$, making flips on odd sites energetically favorable. Thus, the relevant low-energy physics is governed by the fluctuations of odd sites, while even sites act as a rigid background.

\textbf{Higher-Order Expansion Result:}
Applying the SWT formula, we obtain the effective Hamiltonian governing the odd sites (indexed by $j$) up to the second order in $J/h_z$:
\begin{equation}
\label{eq:heff_full}
H_\text{eff} = \sum_{j \in \text{odd}} \left[ \left(\delta_s - \frac{J^2}{8h_z}\right) \sigma_j^z + J \sigma_j^x \right] + \sum_{j \in \text{odd}} \frac{J^2}{16h_z} \sigma_j^z \sigma_{j+2}^z,
\end{equation}
where $\delta_s = h_s - h_z$ is the distance to the spinodal point.
The first term represents an effective field, the second term is the effective transverse field, and the third term represents a generated nearest-neighbor anti-ferromagnetic Ising interaction between odd sites (mediated by virtual excitations of the even sites).

\textbf{Effective Landau-Zener Model:}
In the immediate vicinity of the spinodal point ($\delta_s \to 0$) and assuming we are in a deeply ordered phase ($J \gg J^2/16h_z$), we can neglect the second-order interaction term and the constant energy shift.
The Hamiltonian simplifies to a sum of independent single-spin Hamiltonians:
\begin{equation}
H_\text{eff} \approx \delta_s \sum_{j \in \text{odd}} \sigma_j^z + J \sum_{j \in \text{odd}} \sigma_j^x.
\end{equation}
Under a linear quench $\delta_s(t) = vt$, this is exactly the Hamiltonian for a set of decoupled two-level systems undergoing Landau-Zener transitions. This mapping explains why the breakdown of metastability in this many-body system can be described by the single-particle Landau-Zener formula, as confirmed by our numerical simulations in the main text.

\section{\label{sec:ppxpp_model_prop}
The Third Example: 1D Next-Nearest Neighbor Tilted Ising Model}

In this section, we extend our analysis to a system with longer-range interactions: the one-dimensional Next-Nearest Neighbor (NNN) tilted Ising model. We demonstrate that the extended interaction range enforces a strong kinetic constraint near the quantum spinodal point, leading to an emergent effective description known as the PPXPP model. We first derive this effective Hamiltonian, then thoroughly characterize the equilibrium properties and critical scaling of the pure PPXPP model, and finally present dynamical simulations to verify the emergent $\mathbb{Z}_3$ criticality in the original NNN Ising model.

\subsection{Emergence of the PPXPP Model at the Spinodal Point}

The Hamiltonian of the 1D NNN tilted Ising model is given by:
\begin{equation}\label{eq: nnn ising model ham}
H = -J_{1}\sum_{j} \sigma_{j}^z\sigma^z_{j+1} - J_{2} \sum_{j}\sigma^z_{j}\sigma^z_{j+2} - h_{x}\sum_{j}\sigma^x_{j} - h_{z}\sum_{j}\sigma^z_{j}.
\end{equation}
We set $J_{1}=1$, $J_2>0$, and consider a weak transverse field $h_x \ll 1$. In the presence of a nonzero longitudinal field $h_{z}$, the ground state is ferromagnetic. 

We apply a linear driving protocol $h_z(t)=vt$, starting from a negative field $h_z<0$ where the system is prepared in the all-down ground state $\ket{\downarrow\dots\downarrow}$. To identify the spinodal point, we consider the energy change associated with a single spin flip operator $\sigma_{j}^x$:
\begin{equation}
\langle H \rangle_{\sigma_{j}^x\ket{\psi} } - \langle H \rangle_{\ket{\psi}} =  4(1+J_{2})-2h_{z} +\mathrm{O}(h_x^2).
\end{equation}
The spinodal point, where this excitation gap vanishes, is located at $h_{z}^{sp}=2(1+J_{2})$. 

Near this point, the class of allowed flip operators possesses an emergent $\mathbb{Z}_{3}$ translational symmetry. Specifically, combinations of flip operators such as $\sigma_{j}^x \sigma_{j+k}^x$ are only energetically permissible for a distance $k \geq 3$. Configurations with $k \leq 2$ result in states with significantly energy difference, $4J_{1}$ (for $k=1$) and $4J_2$ (for $k=2$). These states are suppressed. Thus, the relevant low-energy Hilbert subspace strictly prohibits states containing two up-spins with a distance smaller than 2 (e.g., $\ket{\dots\uparrow\uparrow\dots}$ or $\ket{\dots\uparrow\downarrow\uparrow\dots}$). The allowed constrained states are of the form:
\begin{equation}
\ket{\cdots\uparrow\downarrow\downarrow\uparrow\downarrow\downarrow\cdots}, \ket{\cdots\downarrow\uparrow\downarrow\downarrow\uparrow\downarrow \cdots }, \ket{\cdots\downarrow\downarrow\uparrow\downarrow\downarrow \uparrow\cdots}.
\end{equation}

At the quantum spinodal point, the dynamics within this constrained subspace are exactly described by an effective Hamiltonian known as the PPXPP model~\cite{fendley2004PRB}:
\begin{equation}
\label{eq:ham_ppxpp_eff}
H_\text{eff} =  -h_x \sum_{j}P_{j-2}^{\downarrow} P_{j-1}^{\downarrow}\sigma^x_j P_{j+1}^{\downarrow}P_{j+2}^{\downarrow} - h_z^*\sum_{j}\sigma^z_{j},
\end{equation}
where $h_z^* = h_z - h_z^{sp}$ represents the effective detuning from the spinodal point.

\subsection{Equilibrium Properties and Criticality of the PPXPP Model}

\begin{figure}
	\includegraphics[scale=0.6]{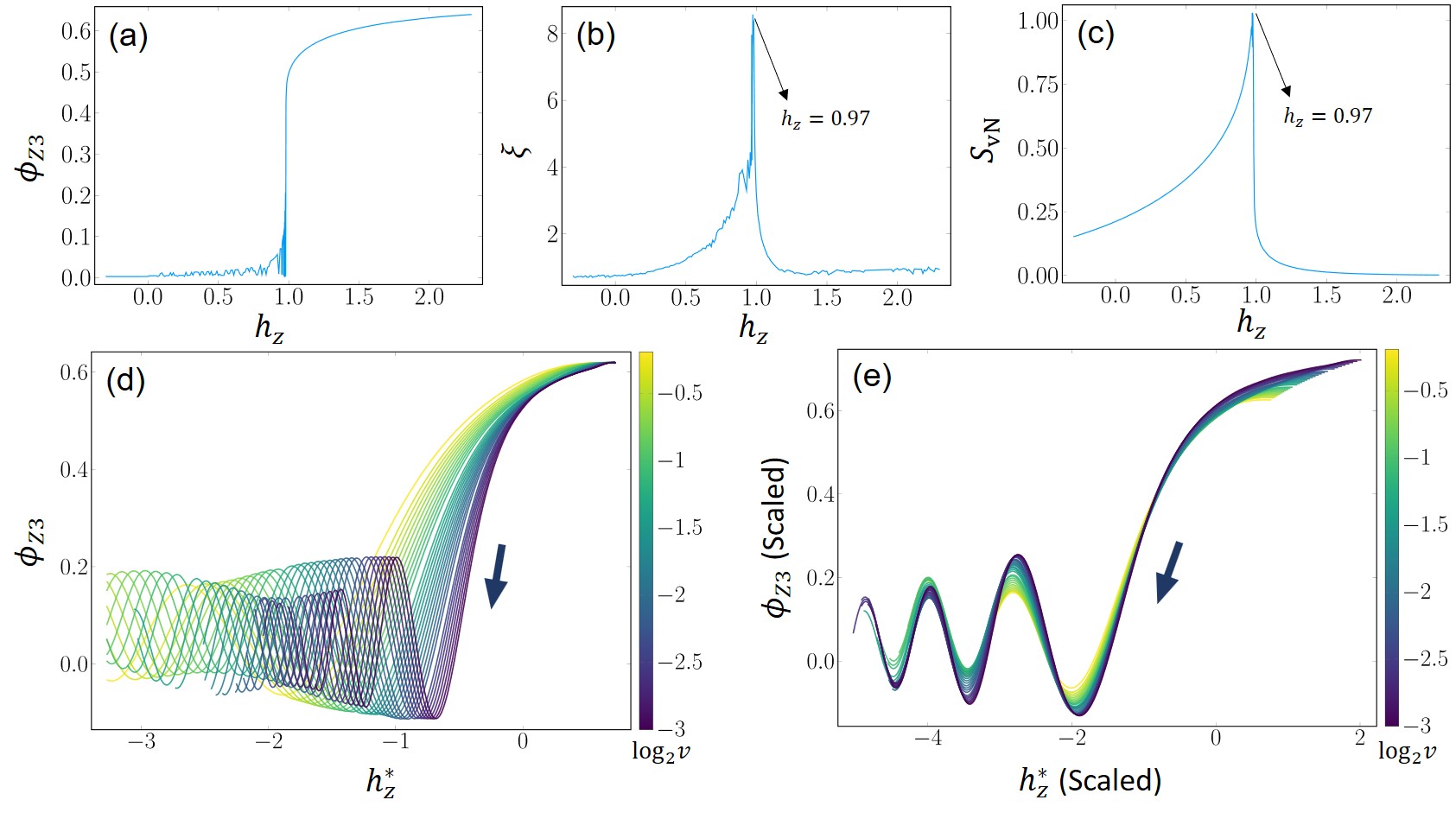}
	\caption{(a)-(c) The equilibrium properties of the PPXPP model. The order parameter (a), the correlation length (b), and the von-Neumann entropy (c) of the ground state as a function of $h_z$. (d), (e) The time evolution of the unscaled (d) and scaled (e) $\mathbb{Z}_3$ order parameter $\phi_{Z3}$. These features clearly show the model exhibits a second-order phase transition  at $h_z=0.97$. The equilibrium properties (a)-(c) are calculated by DMRG under the open boundary condition (OBC) with a size of 352 spins. The evolution (d) and (e) is calculated by TDVP. The scaling relation is $\phi_{Z3}(h_z^*) = v^{0.073}f_{Z3}(h_z^*v^{-0.49})$, $h_z^* = h_z - (h_z)_c = h_z -0.97$.}
	\label{fig:supp_ppxpp}
\end{figure}

To understand the universal dynamics near the spinodal point, we now focus on the properties of the effective PPXPP model itself. This model represents a significant generalization of the standard PXP model, corresponding to the doubly blockaded regime of the interacting hard-core boson model introduced by Fendley, Sengupta, and Sachdev (FSS model)~\cite{fendley2004PRB}:
\begin{equation}
H = - \sum_j (d_j + d_j^\dagger) + U \sum_j n_j + V \sum_j n_j n_{j+2},
\end{equation}
where $d_j^\dagger$ ($d_j$) is the creation (annihilation) operator for a hard-core boson on site $j$ and $n_j = d_j^\dagger d_j$. The Hilbert space is subject to the constraint $n_j n_{j+1} = 0$. In the context of neutral atom quantum simulators, such a regime is experimentally accessible by tuning the parameters of the Rydberg excitation lasers. When the van der Waals interaction strength decays as $V(r) \sim C_6/r^6$, the Rydberg blockade radius $R_b$ can be adjusted to exceed the distance between next-nearest neighbors. In the limit of strong interactions, this effectively prohibits the simultaneous excitation of atoms at distance $|i-j| \le 2$. Consequently, the system is described by hard-core bosons with infinite nearest neighbor ($V_1$) and next-nearest neighbor ($V_2$) repulsion. When $V_2 \to \infty$, the Hamiltonian is given by an effective PPXPP model
\begin{equation}
H_\text{eff} =  -h_x \sum_{j}P_{j-2}^{\downarrow} P_{j-1}^{\downarrow}\sigma^x_j P_{j+1}^{\downarrow}P_{j+2}^{\downarrow} - h_z\sum_{j}\sigma^z_{j},
\end{equation}

According to Ref.~\cite{Giudici2019prb}, the PPXPP model exhibits a rich phase diagram. As the effective field $h_z$ increases, the system transitions from a disordered phase (low density) to a $\mathbb{Z}_3$ ordered phase (high density). This process occurs via two successive transitions separated by an intermediate incommensurate Luttinger liquid (``floating") phase:
\begin{enumerate}
    \item \textbf{Transition 1 ($h_z \approx 0.98$):} A transition from the $\mathbb{Z}_3$ ordered phase to the floating phase. This is a second-order phase transition characterized by critical exponents $z=1.48$, $\nu=0.70$, and $\beta=0.059$.
    \item \textbf{Transition 2 ($h_z \approx 0.96$):} A transition from the floating phase to the disordered phase, belonging to the Berezinskii-Kosterlitz-Thouless (BKT) universality class.
\end{enumerate}

Since the intermediate floating phase is exceedingly narrow, we follow the approach in Ref.~\cite{cheng2023PRB} by treating the critical region as a single effective second-order transition and neglecting the dynamically subtle BKT transition. Our DMRG equilibrium calculations [Figs.~\ref{fig:supp_ppxpp}(a)-(c)] confirm this effective critical point at $(h_z)_c \approx 0.97$, marked by peaks in the correlation length and entanglement entropy. 

The relevant $\mathbb{Z}_3$ order parameter is defined as:
\begin{equation}
\phi_{Z3} = \sum_{j \, \text{mod} \, 3 \, = \, 2} \left( \sigma^z_{j-1} e^{-i \frac{2\pi}{3}} + \sigma^z_{j} + \sigma^z_{j+1} e^{i \frac{2\pi}{3}} \right).
\end{equation}
Based on the extracted critical exponents, the FTS relation for this order parameter under a linear quench $h_z^*(t) = vt$ where $h_z^* = h_z - (h_z)_c$ is given by:
\begin{align}
    \phi_{Z3}(h_z^*) = v^{\beta/(\nu r)}f(h_z^*v^{-1/(\nu r)}) \approx v^{0.073}f(h_z^*v^{-0.49}),
\label{eq:scale_ppxpp}
\end{align}
where $r \equiv z + 1/\nu$. As demonstrated in Fig.~\ref{fig:supp_ppxpp}(e), our TDVP simulations for the pure PPXPP model show that the rescaled order-parameter curves for different quench velocities collapse well onto a single curve, validating the FTS framework for this effective model.

The FTS relation for the correlation length is
\begin{equation}
\xi(h_z^*) =v^{-{1}/{r}} f(h_z^* v^{-{1}/{(\nu r)}}) \approx v^{-0.34}f(h_z^*v^{-0.49}),
\end{equation}
which is particularly useful when analyzing the original NNN Ising model, where the $\mathbb{Z}_3$ order parameter remains zero during the transition from the disordered to the ordered state.

\begin{figure*}[t]
\includegraphics[scale=1.0]{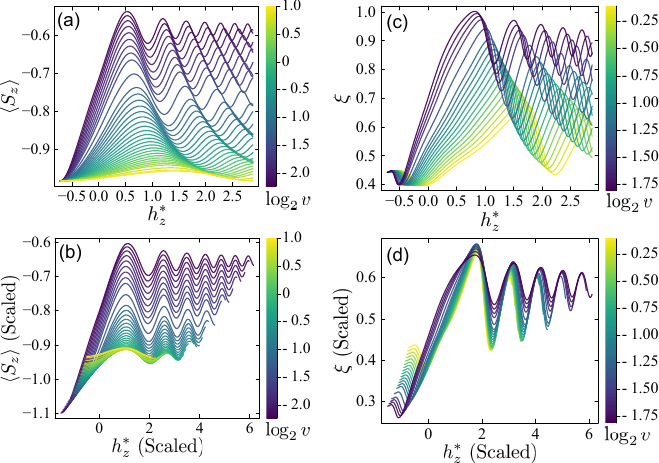}\caption{Time evolution of the 1D NNN tilted Ising model for different ramping rates $v$. Panels (a) and (b) show the unscaled and scaled magnetization $\langle S_z \rangle$ of the NNN tilted Ising model, respectively. Panels (c) and (d) display the unscaled and scaled correlation length $\xi$. Since $\langle S_z \rangle$ is not the true order parameter of the effective PPXPP Hamiltonian, it does not obey a strict scaling ansatz and only the peak positions collapse onto the same scaled field. In contrast, the correlation length exhibits a good scaling collapse, as shown in panels (c)-(d).
\label{fig:nexttitledIsiing}}
\end{figure*}

\subsection{Dynamical Simulation in the NNN Ising Model}

In this section,we finally perform the dynamical simulations of the full 1D NNN tilted Ising model (Eq.~(\ref{eq: nnn ising model ham})) to verify the emergence of criticality near its spinodal point. 

Because the complex $\mathbb{Z}_3$ order parameter $\phi_{Z3}$ in the original NNN Ising model is always zero, we instead calculate the standard longitudinal magnetization $\langle S_z \rangle = \frac{1}{N}\sum_j \langle \sigma^z_j \rangle$ for both the original model and the effective PPXPP model for comparison.

As shown in Fig.~\ref{fig:nexttitledIsiing}(a) and (c), the unscaled magnetization dynamics for both models display distinct peaks. Crucially, when we rescale the axes according to the theoretically derived combinations $1/(\nu r) \approx 0.49$ and $\beta/(\nu r) \approx 0.073$, the peaks of the magnetization for different driving velocities collapse onto the same position in the scaled field (Fig.~\ref{fig:nexttitledIsiing}(b) and (d)). This data collapse, alongside the scaling of the correlation length $\xi$ (Fig.~\ref{fig:nexttitledIsiing}(e) and (f)), confirms that the dynamics near the quantum spinodal point of the NNN Ising model are indeed governed by an emergent second-order quantum phase transition with $\mathbb{Z}_3$ critical exponents. The small discrepancies between the models arise from weak dynamical leakage out of the constrained subspace in the full NNN model.

\bibliographystyle{apsrev4-2}
\bibliography{ref}